\documentclass[%
 reprint,
 amsmath,amssymb,
 aps,prx
]{revtex4-2}

\usepackage{graphicx}% Include figure files
\usepackage{dcolumn}% Align table columns on decimal point
\usepackage{bm}% bold math
\usepackage[utf8]{inputenc}
\DeclareMathOperator{\Tr}{Tr}
\usepackage{epstopdf}
\usepackage{xcolor}

\begin{document}

%\preprint{APS/123-QED}

\title{Active nematics on flat surfaces: from droplet motility and scission to active wetting}% Force line breaks with \\

\author{Rodrigo C. V. Coelho$^{1,2}$}
\email{rcvcoelho@fc.ul.pt}
\author{Hélio R. J. C. Figueiredo$^{1,2}$}% 
\author{Margarida M. Telo da Gama$^{1,2}$}% 
  \affiliation{$^1$Centro de Física Teórica e Computacional, Faculdade de Ciências, Universidade de Lisboa, 1749-016 Lisboa, Portugal.}%Lines break automatically or can be forced with \\
 \affiliation{$^2$Departamento de Física, Faculdade de Ciências,
Universidade de Lisboa, P-1749-016 Lisboa, Portugal.}


\begin{abstract}
We consider the dynamics of active nematics droplets on flat surfaces, based on the continuum hydrodynamic theory. We investigate a wide range of dynamical regimes as a function of the activity and droplet size on surfaces characterized by strong anchoring and a range of equilibrium contact angles. The activity was found to control a variety of dynamical regimes, including the self-propulsion of droplets on surfaces, scission, active wetting and droplet evaporation. Furthermore, we found that on a given surface (characterized by the anchoring and the equilibrium contact angle) the dynamical regimes may be controlled by the active capillary number of suspended droplets. We also found that the active nematics concentration of the droplets varies with the activity, affecting the wetting behaviour weakly but ultimately driving droplet evaporation. Our analysis provides a global description of a wide range of dynamical regimes reported for active nematics droplets and suggests a unified description of droplets on surfaces. We discuss the key role of the finite size of the droplet and comment on the suppression of these regimes in the infinite size limit, where the active nematics is turbulent at any degree of activity.   
\end{abstract}

\maketitle

%\tableofcontents

%%%MAIN TEXT%%%%

\section{Introduction}

Since their discovery, liquid crystals (LC) provided a fertile ground for the development of new theoretical methods and applications. While the direction of molecular alignment in nematics, the director \textbf{n}, is arbitrary in the bulk, in the presence of surfaces and interfaces it selects a particular direction, which is known as surface anchoring. Typical anchorings include homeotropic (perpendicular to the surface), random planar (random in the surface plane) and planar (along one direction in the surface plane). 

In fact, nematic wetting of flat surfaces was predicted in the framework of the Landau–de Gennes (LdG) functional~\cite{PhysRevLett.37.1059, PhysRevLett.55.2907, sluckin1986fluid}, and then observed experimentally~\cite{doi:10.1080/00268948308072027, PhysRevLett.62.1860} shortly before the wetting transition proper was discovered by John Cahn, based on the Landau-Ginzburg free-energy functional for systems with conserved scalar order parameters, such as fluids and fluid mixtures~\cite{doi:10.1063/1.434402, sullivan1986wetting, dietrich1988phase, RevModPhys.81.739}.  

At the nematic–isotropic (NI) coexistence, the quantity of interest is the nematic orientational order parameter \textbf{Q}, a traceless symmetric tensor (to be defined below), as the densities of the coexisting liquid phases are very similar and their spatial variation may be safely neglected. In order to describe nematic wetting, however, we have to consider the anchoring at the surface and at the NI interface. In the simplest case, both favour homeotropic anchoring and the LdG functional reduces to the Landau–Ginzburg free-energy functional of a subcritical fluid at a flat surface (see~\cite{doi:10.1080/00268976.2010.542780} for the mapping and a brief summary of results for different anchorings).

Recently, LCs received renewed attention in a different context, as the LdG free energy turned out to provide a (well understood) starting point for the continuum hydrodynamic theory of active nematics~\cite{PhysRevLett.89.058101}, a new class of soft matter systems characterized by the input of energy at the particle level and its conversion into directed motion~\cite{Gompper_2020, RevModPhys.85.1143}. In active nematics the long-range orientational order in the bulk, i.e., the alignment of the particles, is unstable to (bend or splay) distortions for any degree of activity and leads to a globally disordered state characterized by spontaneous chaotic flows known as active turbulence. This spontaneous flow state exhibits strong vorticity and motile topological defects which are continually created and destroyed. Active turbulence is one of the most striking, if not the most striking, collective behaviour of active matter and it is a current hot topic of research~\cite{doi:10.1146/annurev-conmatphys-082321-035957}

Control of these spontaneous chaotic flows, however, is required for most practical applications. Confinement of active nematics stabilizes the chaotic flow at low activities and the active turbulent state is often preceded by non-steady states at intermediate activities~\cite{Doostmohammadi2018, marenduzzo07steady, C9SM00859D}. It has been shown that some of these systems 
are amenable to controlled macroscopic directed flow, which is key to their use in applications~\cite{doi:10.1126/science.aal1979}.

A second line of research has focused on droplets of active nematics. In addition to controlling the directed motion of self-propelled motion, active nematics droplets provide a model for some biophysical processes. Recent work suggests that the motility, morphological changes and scission of active nematics droplets share mechanisms with similar processes observed in living cells. This spurred the investigation of 2D and 3D active nematics droplets, based on the continuum hydrodynamic theory of active nematics~\cite{PhysRevLett.112.147802, PhysRevX.11.021001}.

Indeed, topological defects in active nematics confined in spherical geometries were proposed to drive the onset of self-propulsion of droplets, through surface induced distortions of the director field which in turn drive flow instabilities~\cite{978-1-83916-229-9}. In 2D systems, the nematics hydrodynamic equations exhibit an instability for activities above a threshold that depends on $R^{-1}$. This argument suggests that self-propulsion results from the interplay of activity and elasticity of droplets with strong surface anchoring. Furthermore it was shown that the suspended droplet dynamical regimes in 2D are controlled by a single physical parameter, namely, an active capillary number~\cite{PhysRevLett.112.147802}. For a review of recent work including results for 3D droplets see~\cite{978-1-83916-229-9}. 

Much less attention has been given to the behaviour of active droplets on surfaces, despite the fact that a range of technological applications and biological processes, such as development and regeneration in tissue morphology, are driven by surface  dynamical instabilities. In analogy with the behavior of passive fluids, some of these transitions have been interpreted as wetting transitions~\cite{doi:10.1126/science.1226418}. This analogy was questioned because the active cellular and the cellular-surface  interactions that drive tissue wetting were not identified or measured. In a recent work that combines experiments and theory it was shown that the transition between 2D epithelial monolayers and 3D aggregates can be understood as an active wetting transition~\cite{PrezGonzlez2018}. Furthermore, the role of an intrinsic lengthscale that controls active wetting was revealed. The latter is absent in passive wetting and it was proposed as one of the distinguishing features of active wetting transitions~\cite{PrezGonzlez2018}.

More recently, a combination of experimental and theoretical work, addressed the question of how mechanical activity shapes the interfaces that separate an active from a passive fluid. In particular, the authors reported that when in contact with a solid surface, the active-passive interface exhibits a non-equilibrium wetting transition, and identified an active interfacial tension that controls the contact angle defined through Young's equation~\cite{doi:10.1126/science.abo5423}. 

At equilibrium, the mechanical interfacial tension and the interfacial free energy are identical. In active fluids this is no longer the case~\cite{PhysRevLett.127.068001,PhysRevLett.115.188302} and active forces may contribute in various ways to the interfacial tensions and drive the active wetting transition. However, it is not clear how to relate the interfacial active stresses considered in~\cite{PrezGonzlez2018, doi:10.1126/science.abo5423} not least because the first is a dry system while in the latter hydrodynamics appears to play a significant role. Furthermore, there are several mechanisms through which the activity may change the wetting phase behaviour, including a shift in the active-passive bulk phase diagram~\cite{doi:10.1146/annurev-conmatphys-031214-014710,TjhungPRX2018,PhysRevLett.129.268002}. 

Central to the work reported here, is the role of the active capillary number introduced in~\cite{PhysRevLett.112.147802}, which controls the dynamics of suspended active nematics droplets, but has not been investigated in the context of active wetting, i.e., when active droplets are deposited on a flat surface. 

Motivated by these fundamental questions we consider the dynamics of active nematics droplets on flat surfaces, based on the continuum hydrodynamic theory. We investigate a wide range of dynamical regimes for droplets with different activities and sizes on surfaces with different anchorings and equilibrium contact angles. The activity was found to drive a variety of dynamical regimes, including self-propulsion of droplets on surfaces, scission, active wetting and droplet evaporation. Furthermore we found that these regimes on a particular surface are controlled by the active capillary number introduced for suspended droplets. We also found that the nematic order parameter in the droplets varies with the activity, affecting weakly the wetting behaviour but ultimately driving droplet evaporation. Our analysis provides a unified overview of the striking dynamical regimes of active nematics droplets on flat surfaces. 

The article is arranged as follows. We start with a brief description of the hydrodynamic model followed by an analytic estimate of the interfacial tension of active nematics, which controls the contact angle through Young's equation. We also discuss the active capillary number that measures the relative importance of the active forces on the droplet and the interfacial tension. Then we present our results for surfaces, which are characterized by different types of (strong) anchoring and thermodynamic contact angles. We consider one type of activity and investigate the dynamical regimes of extensile active nematics droplets by solving the hydrodynamic equations numerically. The effect of droplet size is also analysed in order to confirm the hypothesis that the active capillary number $Ca_\alpha$ encodes the properties of the droplets in a single physical parameter. Particular attention is given to the active wetting transition on planar surfaces and to the transition from linear to chaotic motion on homeotropic surfaces. We end with a summary and a discussion of the results, with emphasis on the active wetting transitions reported in~\cite{doi:10.1126/science.abo5423} and~\cite{PrezGonzlez2018}. 

\section{Theory and methods}

\subsection{Hydrodynamic equations}
\label{hydrodynamic-sec}

For uniaxial nematics, the order is described by the director field $n_\alpha$, which is the average direction of alignment of the particles, and the scalar order parameter $S$, which measures the degree of alignment. These two fields are combined in the tensor order parameter, $Q_{\alpha \beta} = S(n_\alpha n_\beta - \delta_{\alpha \beta}/3)$, which is traceless and symmetric. The equilibrium state of the system is given by the minimum of the Landau-de Gennes free energy $\mathcal{F} = \int_V \,d^3 r\, f_{LdG}$, where the energy density is given by:
\begin{align}
 &f_{LdG}(\gamma) = \frac{A_0}{2}\left( 1- \frac{\gamma}{3} \right) Q_{\alpha \beta}^2 - \frac{A_0\gamma}{3} Q_{\alpha \beta} Q_{\beta \gamma} Q_{\gamma \alpha}  \nonumber \\
 &+ \frac{A_0\gamma}{4} (Q_{\alpha \beta} Q_{\alpha \beta} )^2  + \frac{L}{2} (\partial _\gamma Q_{\alpha \beta})^2 .
 \label{fldg-eq}
\end{align}
Here $A_0$ is a positive constant that sets the energy of the nematic, $L$ is a positive elastic constant that penalizes distortions in $Q_{\alpha\beta}$ and $\gamma$ is the ordering field, e.g., temperature in thermotropic and a concentration related parameter in lyotropic liquid crystals~\cite{doi1988theory, beris1994thermodynamics}. The nematic (N) and isotropic (I) phases coexist when $\gamma=2.7$, where the free energy of the nematic, with a scalar order parameter $S_N=1/3$, is zero. This NI transition is (weakly) first-order, as observed in the experiments, driven by the presence of the cubic term in the free-energy density. For simplicity, we use the single-elastic constant approximation, i.e., we neglect the LC elastic anisotropy~\cite{Coelho2021ptrsa}. 
  
The LdG model does not conserve the N order parameter and a multicomponent model that considers two immiscible conserved fluids, with the concentration given by a scalar field $\phi$: $\phi=+\phi_0$ for the nematic and $\phi=-\phi_0$ for the isotropic fluid, with $\phi_0$ the absolute value of $\phi$ at coexistence, is often employed for lyotropic LCs. The latter are mixtures of conserved nematogen particles in isotropic fluids. We consider the mixture deep in the two-phase region. The free energy density reads $f = f_\phi + f_{LdG}(\gamma(\phi))$, where the contribution from the concentration field is:
\begin{align}
f_\phi&=\frac{a}{4}(\phi^2-1)^2  + \frac{K}{2} \left(\partial_\gamma \phi \right)^2,
\label{energy-mc-eq}
\end{align}
where $a$ is a positive constant that sets the energy related to the $\phi$ field, and $K$ is a positive elastic constant, which penalizes inhomogeneities in $\phi$. The concentration at coexistence is $\phi_0= \pm 1$ and the ordering field $\gamma$ is a function of $\phi$ that sets the coefficient of the quadratic term of the LdG free energy density, Eq.~\eqref{fldg-eq}, around its value at NI coexistence: $\gamma(\phi) = \gamma_0 + \gamma_s(\phi+1)/2$, with $\gamma_0$ being the minimum value of $\gamma$ and $\gamma_s$ being the difference between the values of $\gamma$ in the nematic and isotropic components. The linear dependence of $\gamma(\phi)$ accounts for the increase in the nematic order with the concentration of nematogen particles, which increases with $\phi$. We set $\gamma_0=2.6$ and $\gamma_s=0.2$ which implies that the order is nematic when $\phi=1$ and isotropic when $\phi=-1$. Recall that the LdG free energy exhibits a first order NI transition at $\gamma = 2.7$, which is the value taken by $\gamma$ at $\phi=0$ in the middle of the diffusive interfacial region.

The time evolution of the nematic is governed by the Beris-Edwards equation~\cite{beris1994thermodynamics}, the continuity, the Navier-Stokes equation~\cite{beris1994thermodynamics, landau1987} and the Cahn-Hilliard equation, respectively:
\begin{align}
  &\partial _t Q_{\alpha \beta} + u _\gamma \partial _\gamma Q_{\alpha \beta} - S_{\alpha \beta} = \Gamma H_{\alpha\beta} , \label{beris-edwards-eq} \\
  &\partial _\beta u_\beta = 0, \label{continuity-eq}\\
&\rho\partial_t  u_\alpha + \rho u_\beta \partial _\beta   u_\alpha  =  \partial_\beta [2\eta D_{\alpha\beta}  + \sigma^{\text n}_{\alpha\beta} -\zeta Q_{\alpha\beta} ], \label{navier-stokes-eq}\\
 &\partial_t \phi + \partial_\beta (\phi u_\beta) = M \nabla^2 \mu, 
 \label{cahn-hilliard-eq}
\end{align}
where $D_{\alpha\beta} = (\partial _\beta u_\alpha + \partial _\alpha u_\beta )/2$ is the shear rate.
Equation~\eqref{beris-edwards-eq} describes the evolution of the order parameter $Q_{\alpha\beta}$, Eqs.~\eqref{continuity-eq} and \eqref{navier-stokes-eq} describe the dynamics of the velocity field $u_\alpha$, while Eq.~\eqref{cahn-hilliard-eq} describes the evolution of the field $\phi$. Here $\Gamma$ is the system dependent rotational diffusivity, $\rho$ is the density, $\eta$ is the shear-viscosity and $M$ is the mobility constant that controls the diffusion of the concentration field. The last term in Eq.~\eqref{navier-stokes-eq} is the active stress, which corresponds to a force dipole density, with $\zeta$ the activity parameter being positive for extensile stresses (systems of pushers) and negative for contractile ones (systems of pullers)~\cite{PhysRevLett.89.058101}. Thus gradients in $Q$ produce a flow field, which is the source of the hydrodynamic instabilities reported in bulk and confined active nematics. The co-rotational term is as follows: 
\begin{align}
 &S_{\alpha \beta} = ( \xi D_{\alpha \gamma} + W_{\alpha \gamma})\left(Q_{\beta\gamma} + \frac{\delta_{\beta\gamma}}{3} \right) 
 + \left( Q_{\alpha\gamma}+\frac{\delta_{\alpha\gamma}}{3} \right) \nonumber \\& 
 \cdot(\xi D_{\gamma\beta}-W_{\gamma\beta}) - 2\xi\left( Q_{\alpha\beta}+\frac{\delta_{\alpha\beta}}{3}  \right)(Q_{\gamma\epsilon} \partial _\gamma u_\epsilon), 
 \label{corrotational-eq}
\end{align}
where $W_{\alpha\beta}= (\partial _\beta u_\alpha - \partial _\alpha u_\beta )/2$ is the vorticity
and $\xi$ is the flow alignment parameter, which characterizes the relative importance of the shear rate and the vorticity in the flow alignment of the particles.
The molecular field $H_{\alpha\beta}$ describes the relaxation of the order parameter towards equilibrium:  
\begin{align}
 H_{\alpha\beta} = -\frac{\delta \mathcal{F}}{\delta Q_{\alpha\beta}} + \frac{\delta_{\alpha\beta}}{3} \Tr \left( \frac{\delta \mathcal{F}}{\delta Q_{\gamma \epsilon}} \right).
\end{align}
The passive nematic stress tensor is~\cite{beris1994thermodynamics}:
\begin{align} 
 \sigma_{\alpha\beta}^{\text n} =& -P_0 \delta_{\alpha\beta} + 2\xi \left( Q_{\alpha\beta} +\frac{\delta_{\alpha\beta}}{3} \right)Q_{\gamma\epsilon}H_{\gamma\epsilon} \nonumber \\ 
 & - \xi H_{\alpha\gamma} \left( Q_{\gamma\beta}+\frac{\delta_{\gamma\beta}}{3} \right) - \xi \left( Q_{\alpha\gamma} +\frac{\delta_{\alpha\gamma}}{3} \right) H_{\gamma \beta} \nonumber \\ 
 & +\sigma^{\text{s}}_{\alpha\beta}  + Q_{\alpha\gamma}H_{\gamma\beta} - H_{\alpha\gamma}Q_{\gamma\beta} ,
 \label{passive-pressure-eq}
\end{align}
where $P_0$ is the isotropic pressure.
The chemical potential is given by:
\begin{eqnarray}
 \mu = \frac{\partial f}{\partial \phi}-\partial _\gamma \left[    \frac{\partial f}{\partial \left( \partial_\gamma \phi \right)}\right].
\end{eqnarray}
 In Eq.~\eqref{passive-pressure-eq}, the term $\sigma^s_{\alpha\beta}$ is:
\begin{align}
 \sigma^s_{\alpha\beta} &= \left( f-\mu \frac{(\phi+1) }{2}\right) \delta_{\alpha\beta} - \frac{\delta \mathcal{F}}{\delta \left(  \partial_\beta \phi \right)} \partial_\alpha \phi \nonumber \\
 &- \frac{\delta \mathcal{F}}{\delta \left(   \partial_\beta Q_{\gamma\nu}\right)}\partial_\alpha Q_{\gamma \nu}
\end{align}
The results and parameters are expressed in simulation units: the lattice spacing corresponding to the spatial step is $\Delta x=1$, the time step is $\Delta t = 1$ and the reference density is $\rho_{\text{ref}}=1$. The conversion to physical units is given either by setting appropriate values to these three quantities or by comparing non-dimensional numbers as the active Capillary number that will be described. 

The simulations were performed in a 2D domain of width \(L_X = 256\) and height \(L_Y = 128\) with periodic boundary condition in the $x$-direction. There are two identical surfaces: one at the bottom and another at the top, both with no-slip boundary conditions for the velocity field, with the same equilibrium contact angle and with the same director alignment (surface anchoring). The distance between the two surfaces is larger than the other relevant length scales: droplet radius, active length, vortex size and nematic correlation length. A nematic droplet with radius \(R = 22.4\) (except where stated otherwise) is initialized at $x=L_X/2$ and $y=R$ with uniform alignment, i.e., it starts as a circular droplet touching the surface and then spreads. The simulations run up to $t=3\times 10^6$ iterations. 
The wetting boundary conditions follow Ref.~\cite{VANDERSMAN20132751}. We use half-way bounce-back~\cite{kruger2016lattice} conditions for the populations corresponding to the Navier-Stokes equation, which results in no-slip conditions at the surface. 
The system of differential equations is solved using a hybrid method with the same spatial discretization: Eq.~\eqref{beris-edwards-eq} is solved using finite-differences and Eqs.~\eqref{continuity-eq}, \eqref{navier-stokes-eq} and Eq.~\eqref{cahn-hilliard-eq} are recovered in the macroscopic limit with the lattice-Boltzmann method. The numerical method is similar to those used in Refs.~\cite{C9SM00859D, refId0,C9SM02306B} and, for simplicity, we use the single relaxation time approximation in the Boltzmann equation~\cite{succi2018lattice, kruger2016lattice}. 
The parameters used in the simulations, except when stated otherwise, are: $L=0.04$, $\xi=0.7$ (flow aligning), $A_0=0.1$, $\rho=10$, $\tau=1.0$ (or, kinematic viscosity $\nu=(\tau-1/2)/3=0.17$), $\Gamma=0.34$, $K=0.08$, $a=0.05$ and $M=0.5$.

The nematic correlation length is the square root of the ratio of the quadratic terms of the LdG free energy density, Eq.~\eqref{fldg-eq}. Assuming that the director is uniform, Eq.~\eqref{fldg-eq} simplifies (see the Appendix) and in the single elastic constant approximation the perpendicular and tangential correlation lengths are equal and given by
\begin{align}
\ell_N=\sqrt{\frac{L}{A_0 \left( 1- \frac{\gamma}{3} \right)}}
\label{nematic-correlationlength-eq}
\end{align}
We note that this correlation length depends on the value of the ordering field $\gamma$ that varies throughout the interface. At the NI transition, right in the middle of the diffusive interface ($\phi=0$) the nematic correlation length is $\ell_N=\sqrt{\frac{10 L}{A_0}} \approx 2$, for the parameters given in the Appendix, setting the (free) NI interfacial width $2\ell_N \approx 4$ (see the Appendix).

Similarly, the correlation length associated with the concentration field $\phi$ is 
\begin{align}
\ell_\phi = \sqrt{K/a}
\label{phi-correlationlength-eq}
\end{align}
For the parameters given in the Appendix $\ell_\phi\approx 1.26$ setting the equilibrium interfacial width of the concentration field, $\sqrt{2}\ell_\phi \approx 1.78$. We note that both correlation lengths and interfacial widths are similar and larger than the lattice spacing ($\Delta x =1$). 

In the Appendix we compare the orientational order parameter $S$ and the concentration $\phi$ profiles of active nematics droplets obtained numerically using the hydrodynamic equations, with the approximate analytical expressions for the flat passive interfaces  discussed above and find excellent agreement at low activities for large droplets (see the Appendix for the details and Figs.~\ref{s-phi-profile-fig} and~\ref{teo-an-fig} for the numerical and analytic profiles). 

\subsection{Interfacial tension and wetting}
\label{sigma-sec}

The interfacial tension between the passive nematic and the isotropic phase at coexistence is the integral of the excess free energy density across the interface: $\Sigma = \int_{-\infty}^\infty f dx$. This has two contributions: one from $f_\phi$ and the other from $f_{LdG}$. We note that the free energy of the coexisting phases is zero (by construction) and assume that the concentration and tensor order parameter fields, $\phi$ and $Q_{\alpha\beta}$, are independent in order to estimate the contributions to the interfacial tension of a flat interface from the spatial variation of each of the two fields. For the concentration field $\phi$ we obtain the well known result, 
\begin{align}
\Sigma_\phi=\int_{-\infty}^\infty f_\phi dx= \sqrt{\frac{8Ka}{9}} \phi_0^2, 
\label{sigma-phi-eq}
\end{align} 
where $x$ is the coordinate normal to the interface and we used the equilibrium concentration profile obtained by minimizing the free energy density, 
\begin{align}
\phi(x)=\phi_0\tanh\left(\frac{x}{\sqrt{2}\ell_\phi}\right),
\label{phi-eq-profile}
\end{align}
with $\phi_0=\pm 1$ the concentrations at coexistence. Note that the interfacial tension depends on the values of $\phi$ at coexistence, which may change with the activity as we will discuss in the next sections. A similar calculation for the $Q_{\alpha\beta}$ field, yields
\begin{align}
\Sigma_N=\int_{-\infty}^\infty f_{LdG} dx= \frac{\sqrt{ A_0 L }} {81\sqrt{10} },
\label{sigma-Q-eq}
\end{align}
at the NI coexistence ($\gamma=2.7$). We assumed that the director field is uniform and homeotropic and thus the LdG free energy depends only on the scalar order parameter $S$. The equilibrium profile is obtained by minimizing the free energy density (see the Appendix),
\begin{align}
S=\frac{S_N}{2}\left[\tanh\left(\frac{x}{2\ell_N}\right)+1\right],
\label{S-eq-profile}
\end{align}
where $S_N$ is the value of the nematic scalar order parameter at NI coexistence. We assumed the coefficients of the LdG free energy are independent of $\phi$, and set them to those at the NI transition. These are the coefficients of the LdG free energy at the centre of the diffuse interface, $\phi=0$, and we expect Eq.~\eqref{sigma-Q-eq} to provide a good estimate of the contribution of the LdG free energy Eq.~\eqref{fldg-eq} to the total interfacial tension. For the parameters used here (see the Appendix) we find $\Sigma_\phi/\Sigma_N \approx 241.5$ and thus the contribution of the $\phi$ field to the interfacial tension is clearly dominant. We can therefore neglect the contribution from the $Q_{\alpha\beta}$ field and use $\Sigma \approx \Sigma_\phi= 0.06$. We can argue that the contribution from the orientational order parameter field to the interfacial tension is near-critical and thus much smaller than the contribution from the concentration field. This is corroborated by inspection of the order-parameter profiles for passive and active droplets in Figs.~\ref{s-phi-profile-fig} and~\ref{teo-an-fig}, where the width of the $S$ profile is $\approx 2.8$ times larger than that of $\phi$. We recall that the coefficient of the quadratic term of the LdG free energy density Eq.~\eqref{fldg-eq} varies around its value at the NI transition in the passive system and this implies that the nematic order parameter field varies on longer lengthscales than the concentration field, which is deep in the phase separated regime. 

We now turn our attention from the flat interface to nematic droplets on flat surfaces. At the thermodynamic level, a passive liquid droplet deposited on a flat solid surface will form a spherical droplet, defining the equilibrium contact angle of the liquid with the surface. The cornerstone of wetting phenomena, a force balance known as Young’s equation, relates the contact angle $\theta_c$ with the interfacial tensions of the solid–vapour $\Sigma_{sv}$, solid–liquid $\Sigma_{sl}$, and liquid–vapour $\Sigma$ interfaces: 
\begin{align}
\cos(\theta_c) = \frac{\Sigma_{sv} -  \Sigma_{sl}} { \Sigma}.
\label{contatc-angle-eq}
\end{align}
When the contact angle is zero, the liquid spreads to cover the surface and we say that the liquid wets the solid. A wetting transition occurs when the contact angle changes from a finite value to zero, as the temperature or the surface properties vary. The wetting transition of simple fluids on flat surfaces attracted enormous attention a few decades ago and is now well understood~\cite{doi:10.1063/1.434402, sullivan1986wetting, dietrich1988phase, RevModPhys.81.739}. In the simplest case for nematic droplets, both the surface and the NI interface favour homeotropic anchoring, in which case the LdG functional reduces to the Landau–Ginzburg free-energy functional of a subcritical fluid at a flat surface~\cite{doi:10.1080/00268976.2010.542780}.

We define the difference between the interfacial tensions of the surface with the two coexisting fluid phases,  $\Sigma_s = \Sigma_{sv} -  \Sigma_{sl}$ and note that $\Sigma_s > 0$ in the partial wetting regime where $\theta_c < 90^\circ $ and $\Sigma_s < 0$ in the non-wetting or partial drying regime. At $\theta_c = 90^\circ$ the surface wetting is neutral. 
For a system in the partial wetting regime, the threshold for wetting $\theta_c = 0^\circ$ occurs by decreasing the interfacial tension $\Sigma$, an argument used by John Cahn in his seminal work for critical point wetting~\cite{doi:10.1063/1.434402} or by increasing the surface term $\Sigma_s$. For a given solid surface the thermodynamic $\Sigma_s$ is fixed but additional forces will arise driven by the activity. We note that for extensile active nematics the active forces increase $\Sigma_s$ as we will discuss later. For contractile systems the sign of the active forces is reversed. 

\begin{figure}[h]
\centering
\includegraphics[width = \linewidth]{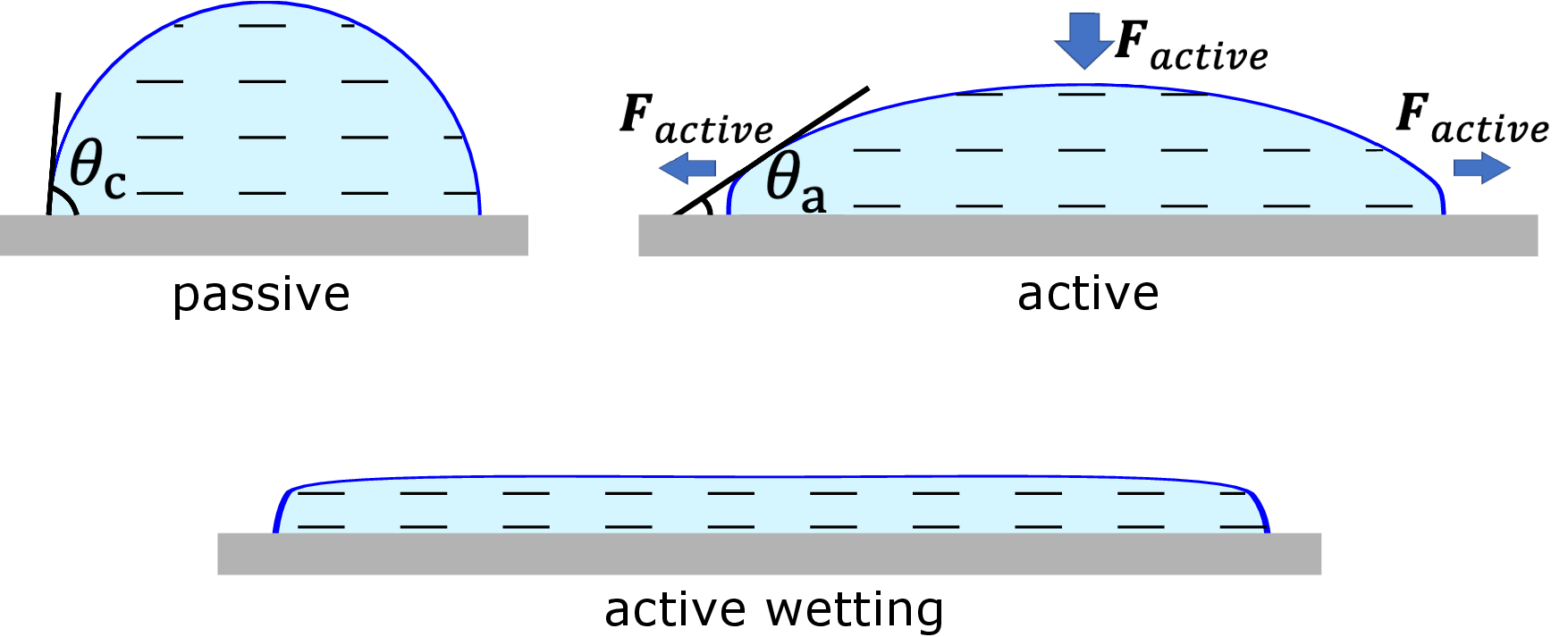}
\caption{Scheme illustrating the shape of a passive droplet (top left), of an extensile active droplet (top right) and of a flat droplet in the active wetting regime (bottom), i.e., with zero apparent contact angle, on a surface with planar anchoring. The active force and the apparent contact angle $\theta_a$ are depicted on the active droplet at the top right.}
\label{fig:scheme}
\end{figure}

The equilibrium contact angle $\theta_c$ of the passive system is controlled by the value of $\phi$ at the surface ($\phi_s$) which is set in the simulations as:
\begin{align}
\cos(\theta_c) = \frac{3}{2}\frac{\phi_s}{\phi_0} \left( 1-\frac{\phi_s^2}{3\phi_0^2}\right).
\label{theta_c}
\end{align}
Note that this equation is obtained using the $\phi$ field and neglects the orientational order parameter contribution to the interfacial tension, which as we have discussed is much smaller. The value of $\phi_0$ at coexistence, however, may change with the activity due to local shearing effects as reported recently~\cite{PhysRevLett.129.268002}. Thus, the contact angle may depend also on the activity through a shift in $\phi _0$. As we will show later this effect is sub-dominant when compared to the effect of the active forces on the droplet in the force balance equation.

\subsection{Active forces: Active capillary number}

We now turn our attention to active nematics droplets on flat surfaces. Bulk active nematics attracted enormous attention thanks to their success in describing novel collective behaviour such as active turbulence. Active nematics droplets were also studied as their self-propulsion is technologically relevant and they provide simple models of biological processes such as cell motility and scission. The active interactions of the particles with each other and with the surrounding medium give rise to active mechanical stresses and flows, which are responsible for these and other phenomena not observed in passive systems. Recently, the spreading of a droplet of epithelial cells on a flat surface was reported and the transition was shown to be driven by the balance of (cell-cell and cell-surface) active stresses, which have been measured. The transition was coined active wetting~\cite{PrezGonzlez2018}.

The hydrodynamic model described in the previous section can include various types of active stress. Some, which we will call elastic active stresses, are coupled to the interfacial square gradients of the order parameters (through the constants $K$ and $L$) and were introduced in the context of scalar order parameters in the active model H~\cite{PhysRevLett.127.068001,PhysRevLett.115.188302}. Another type 
of active stress is linearly coupled to the orientational order parameter $Q_{\alpha \beta}$ and was introduced in the context of the hydrodynamic theory of active nematics in~\cite{PhysRevLett.89.058101}. For simplicity, we consider only the latter type of active stresses, which is known to drive spontaneous flows through the Navier-Stokes Eq.~\eqref{navier-stokes-eq}. This type of active stress corresponds to a force dipole density, with $\zeta$ the strength of the activity being positive for extensile stresses (pushers) and negative for contractile ones (pullers)~\cite{PhysRevLett.89.058101}. A very similar model was used recently to describe the active wetting transition reported experimentally for an active-passive interface in contact with a solid surface~\cite{doi:10.1126/science.abo5423}. A closely related model was also used in~\cite{PhysRevLett.112.147802} to study the dynamics of 2D suspended droplets, where a range of dynamical regimes was reported, and it was shown that the droplet dynamics is controlled by a single physical parameter corresponding to an active variant of the capillary number.

In fluid dynamics, the capillary number (Ca) is a dimensionless quantity that measures the relative effect of viscous drag and interfacial tension forces acting across an interface between a liquid and a gas, or between two immiscible liquids.
A neutrally buoyant droplet placed in a shear flow experiences a strain that scales linearly with the capillary number $Ca = \eta U/\Sigma$ where $U$ is a typical flow velocity and $\eta$ is the fluid viscosity. For active nematics, the typical velocity of the 
flow generated by the defects scales as $U\zeta R/\eta$~\cite{PhysRevLett.110.228101} and Giomi and Simone defined an 
active capillary number as
\begin{equation}
Ca_\alpha = \frac{\zeta R}{\Sigma}.
\end{equation}

We note that $\Sigma$ is the total interfacial tension. As we have discussed above, the contribution to $\Sigma$ from the spatial variation of the orientational order parameter is negligible and the interfacial tension may be approximated by Eq.~\eqref{sigma-phi-eq}. 

The stress tensor includes bulk elastic and active contributions. Assuming that the bulk elastic contribution is much smaller than the active stresses the force balance Eq.~\eqref{navier-stokes-eq} in the Stokes regime implies that $\eta \nabla^2 \mathbf{u} = \zeta \nabla \cdot \mathbf{Q}$. Dimensional analysis, then suggests $\eta U \sim R \zeta$, where $R$ is the characteristic length scale, i.e., the radius of the droplet. Thus, we can extend the active capillary number defined in~\cite{PhysRevLett.112.147802} to active fluids at low velocities with non-singular director distortions.

Finally, we note that the active capillary number may be interpreted as an active Bond number (Bo), a dimensionless number that measures the importance of active (rather than gravitational) forces compared to the interfacial tension in the movement of a liquid front. The active Bond number is then the ratio of the active stress $\zeta R$ to the interfacial tension $\Sigma$. This interpretation is useful as it allows us to use the same number for wet and dry systems where hydrodynamic flows are absent. One important consequence that follows from these numbers is that the size of the droplet, measured by its radius $R$ at equilibrium, amplifies the effect of the activity: increasing the droplet radius at a non-zero activity is equivalent to increasing the activity. The dependence of active wetting on the droplet radius was considered to be a distinguishing feature of the active wetting transition in~\cite{PrezGonzlez2018}. The latter model is substantially different from the hydrodynamic model of the active-passive mixture used here and in~\cite{ doi:10.1126/science.abo5423,PhysRevLett.110.228101} but the scaling of the active stress anticipates the existence an active wetting transition at a threshold activity that depends on the droplet radius. 

\begin{figure*}[t]
\centering
\includegraphics[width = 0.9\linewidth]{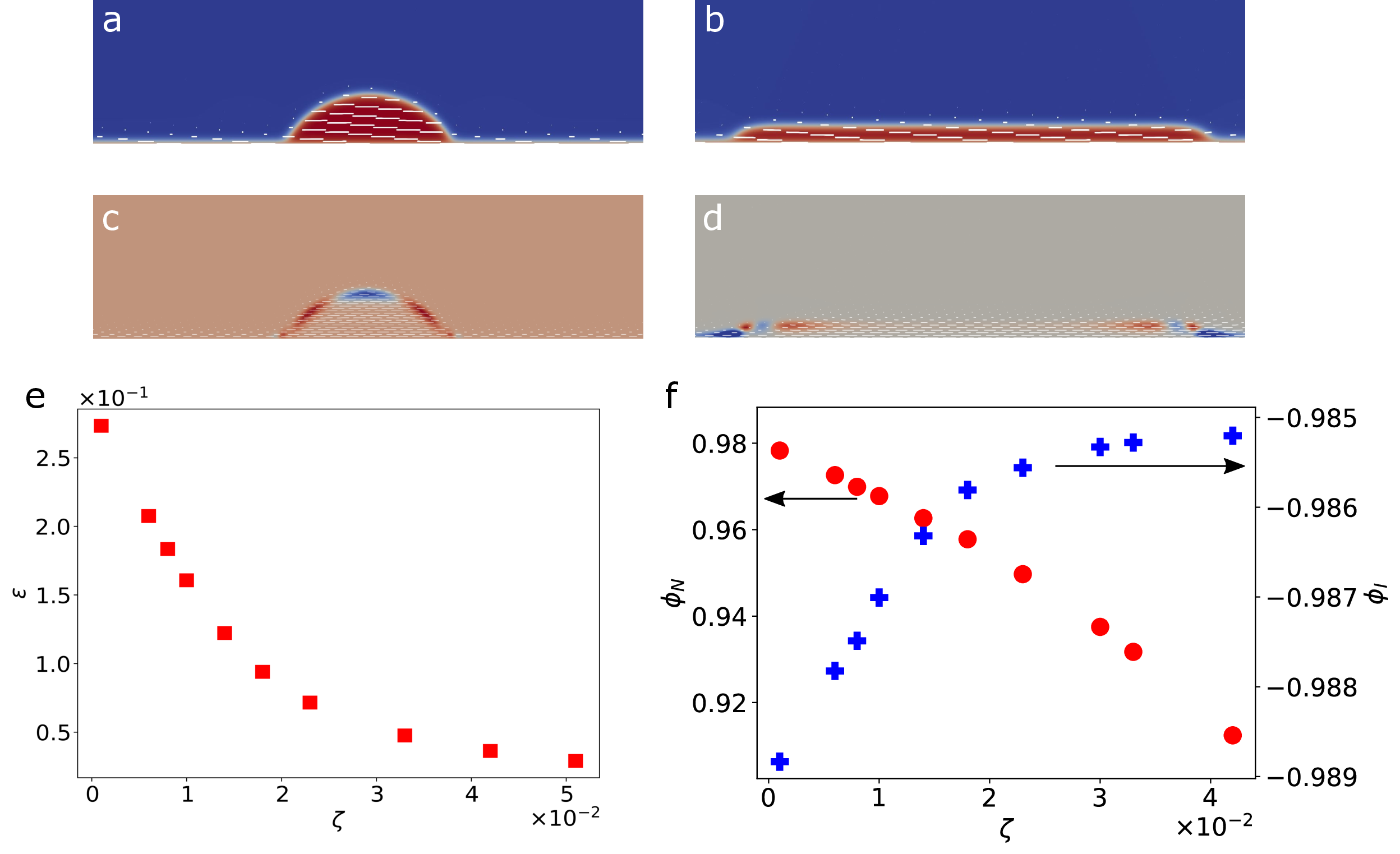}
\caption{Droplet on a surface with planar anchoring, and partial wetting contact angle $\theta_c=60^\circ$. (\textbf{a}) and (\textbf{b}) director field in a droplet with $\zeta=0.001$ and $\zeta=0.05$ respectively.  (\textbf{c}) and (\textbf{d}) Distortion or charge density field for the droplets in (\textbf{a}) and (\textbf{b}) respectively. The red (blue) represents positive (negative) charge density, i.e., bend (splay) distortions. (\textbf{e}) Aspect ratio, \(\epsilon\)=height/width, as a function of the activity, $\zeta$. At large activities $\zeta$ the droplets form wetting films on the surface. (h) Average value of $\phi$ as a function of the activity, $\zeta$, in the nematic ($\phi_N$, on the left, red circles) and isotropic ($\phi_I$, on the right, blue crosses) phases.}
\label{fig:horizontal4}
\end{figure*}

\section{Results}

\subsection{Planar anchoring}
\label{sec:sec:directHoriz}

We start by analysing the behaviour of extensile active nematics droplets on flat surfaces with strong planar anchoring. We consider an equilibrium partial wetting contact angle, $\theta_c=60^\circ$. As we neglect the elastic anisotropy there is no anchoring at the passive NI interface. At the active interface, however, the anchoring is planar for extensile systems with flow aligning particles ($\xi>0$). This is known as active anchoring~\cite{PhysRevLett.113.248303,Coelho2021ptrsa} and its strength increases with the activity.  We find nearly uniform parallel director fields (see Figs.~\ref{fig:horizontal4}(a) and (b)), except in the interfacial region (see Figs.~\ref{fig:horizontal4}(c) and (d)). At low activities the surface anchoring dominates while at high activities the droplet flattens and both the surface and the active anchorings favour parallel alignment, resulting in nearly uniform parallel director fields over a wide range of activities.  

As the activity increases, the droplet spreads increasing the contact area with the surface and lowering the apparent contact angle.  This is quantified by the droplet aspect ratio $\varepsilon=$ height/width plotted in Fig.~\ref{fig:horizontal4}(e). 

The contact angle varies with the activity. The interfacial tension $\Sigma$ which is dominated by the spatial variation of the concentration field scales with the square of the order parameter, $\phi_0^2$, Eq.~\eqref{sigma-phi-eq}. Through a dynamical process known as local shear mixing~\cite{PhysRevLett.129.268002} the affects shifts the coexistence values of the concentration, by reducing them. This has been reported quantitatively for a model similar to ours~\cite{PhysRevLett.129.268002} and was checked explicitly in Fig.~\ref{fig:horizontal4}(f) by calculating the nematic concentration of the droplets at different activities. This shift of the binodal is similar to the shift resulting from the addition of impurities that lower the critical point or increasing the effective temperature of the non-equilibrium system~\cite{PhysRevLett.129.268002}.

For $\theta_c=60^\circ$ and activities that vary from $\zeta=0$ to $0.05$ (static droplet), the measured contact angle decreases by less than $10^\circ$. This may be understood in part using Eq.~\eqref{theta_c}. If the value of $\phi_s$ (concentration at the surface) is fixed and $\phi_0$ changes from $1$ to $0.92$ (see Fig.~\ref{fig:horizontal4}(f)), the equilibrium contact angle changes from $60^\circ$ to $57.4^\circ$. This is smaller than the observed reduction in the contact angle and far too small to drive active wetting. The latter is defined by an apparent contact angle that vanishes as the droplet flattens, see Fig.~\ref{fig:scheme}. 

The flattening of the droplet is driven by extensile active stresses acting on the surface of the droplet, and ultimately responsible for the vanishing of the apparent contact angle, which we define as active wetting. 

The active forces on the surface of the droplet are given by the divergence of the active stress: $F_\alpha^{\text{active}} = -\zeta \partial_\beta Q_{\alpha\beta}$. The projection of this force on the outward normal of the NI interface $\mathbf{m}$ yields the active force perpendicular to the droplet surface. Assuming that the director field is uniform, the forces on the top and on the sides of the droplet are, respectively (see Ref.~\cite{D2SM00988A} for a similar calculation):
\begin{align}
\mathbf{F}^t_\perp = -\frac{\zeta \vert \nabla S \vert}{3}\mathbf{m} \quad \text{and} \quad \mathbf{F}^s_\perp = \frac{2\zeta \vert \nabla S \vert}{3}\mathbf{m}.
\label{active-force-eq}
\end{align}

The extensile active force on the top of the droplet points inwards while it points outwards on the sides. This drives droplet spreading ultimately resulting in a flat wetting film as illustrated in Fig.~\ref{fig:horizontal4}(a) and (b). At low activities, the aspect ratio varies linearly with the activity, but departs from the linear at $\zeta_w\approx 0.015$ in the flat wetting film regime where the apparent contact angle vanishes. This activity can be used to estimate the threshold for the active wetting transition. At the threshold, most of the droplet perimeter is flat. For computational reasons, later on (Sec.~\ref{dropsize-sec}), we define the wetting threshold as the activity when  60\% of the droplet perimeter is flat, which leads to a slightly lower value of $\zeta_w=0.013$. Note that these forces do not act at the contact line and thus lead to an apparent contact angle $\theta_a$, which differs from the contact angle $\theta_c$. This threshold is independent from the size of the simulation box and allows us to define the active wetting transition.

The active force in Eq.~\eqref{active-force-eq} assumes a uniform director field but there are small distortions close to the interface due to active anchoring. These distortions may be quantified by calculating the charge density field~\cite{PhysRevLett.113.248303, Hardoin2022}:
\begin{align}
q=\frac{1}{\pi} \left( \partial_x Q_{x\alpha} \partial_y Q_{y\alpha} - \partial_x Q_{y\alpha} \partial_y Q_{x\alpha}   \right).
\end{align}
As we will show in the next section, droplet motion occurs in the direction of the positive charge density when the symmetry of the charge distribution is broken. For planar anchoring, this distribution is symmetric, see Figs.~\ref{fig:horizontal4}(c) and (d), and the droplets remain static. 

The mechanism for active wetting may be understood qualitatively as follows. If the equilibrium contact angle is less than $90^\circ$ (partial wetting) the active forces on the sides of the droplet add to $\Sigma_{sl}$ in Young's Eq.~\eqref{contatc-angle-eq} with the opposite sign and increase the difference $\Sigma_s=\Sigma_{sv}-\Sigma_{sl}$, which in turn increases the cosine and decreases the (apparent) contact angle until it vanishes at the active wetting transition. At larger contact angles the droplet deforms and spreads to some extent but the apparent contact angle does not vanish in this range of activities. 

In the non-wetting regime $\cos(\theta_a)$ is negative and larger active forces would be required to promote wetting. We note that the droplet height also increases with the contact angle and the droplet may exhibit other activity driven dynamical regimes that pre-empt active wetting. Indeed, at the neutral contact angle, $\theta_c=90^\circ$, we found that the droplet with $R=22.4$ starts moving with constant velocity at $\zeta\approx 0.03$ before a wetting film is formed. 

An active wetting transition was reported in experiments with an extensile microtubule-kinesin mixture on a surface with planar anchoring~\cite{doi:10.1126/science.abo5423}. Both the structure of the wetting film and the active wetting mechanism are the same as those described here. 

In summary, active extensile forces at the interface of an active droplet on a surface with planar anchoring will oppose the thermodynamic solid-liquid interfacial tension in Eq.~\eqref{contatc-angle-eq}. This increases $\Sigma_s$ resulting in the increase of $\cos(\theta_c)$ and a lower apparent contact angle, which may vanish at an active wetting transition. The active forces scale with the radius of the droplet and will, at fixed activity, drive a wetting transition for sufficiently large droplets if not pre-empted by other dynamical regimes. In the latter case, at a given activity, there is a critical size of the droplet where a wetting transition occurs, something which does not happen in passive systems. The dependence on $R$ or the existence of a critical droplet size at the active wetting transition was reported in a recent work, based on a model with two active forces, as a signature of active wetting~\cite{PrezGonzlez2018}. 

We stress that the arguments discussed above hold far from the active turbulent regime, which will set in at vanishingly small activities for infinitely large droplets. For the droplet considered in this section the film becomes turbulent at $\zeta>0.14$, an activity which is larger that that at the active wetting threshold. The size dependence of the threshold for active wetting will be discussed further in Sec.~\ref{dropsize-sec} and in the Conclusion. 

\subsection{Homeotropic anchoring}
\label{sec:sec:directVert}

\begin{figure}
\centering
\includegraphics[width = \linewidth]{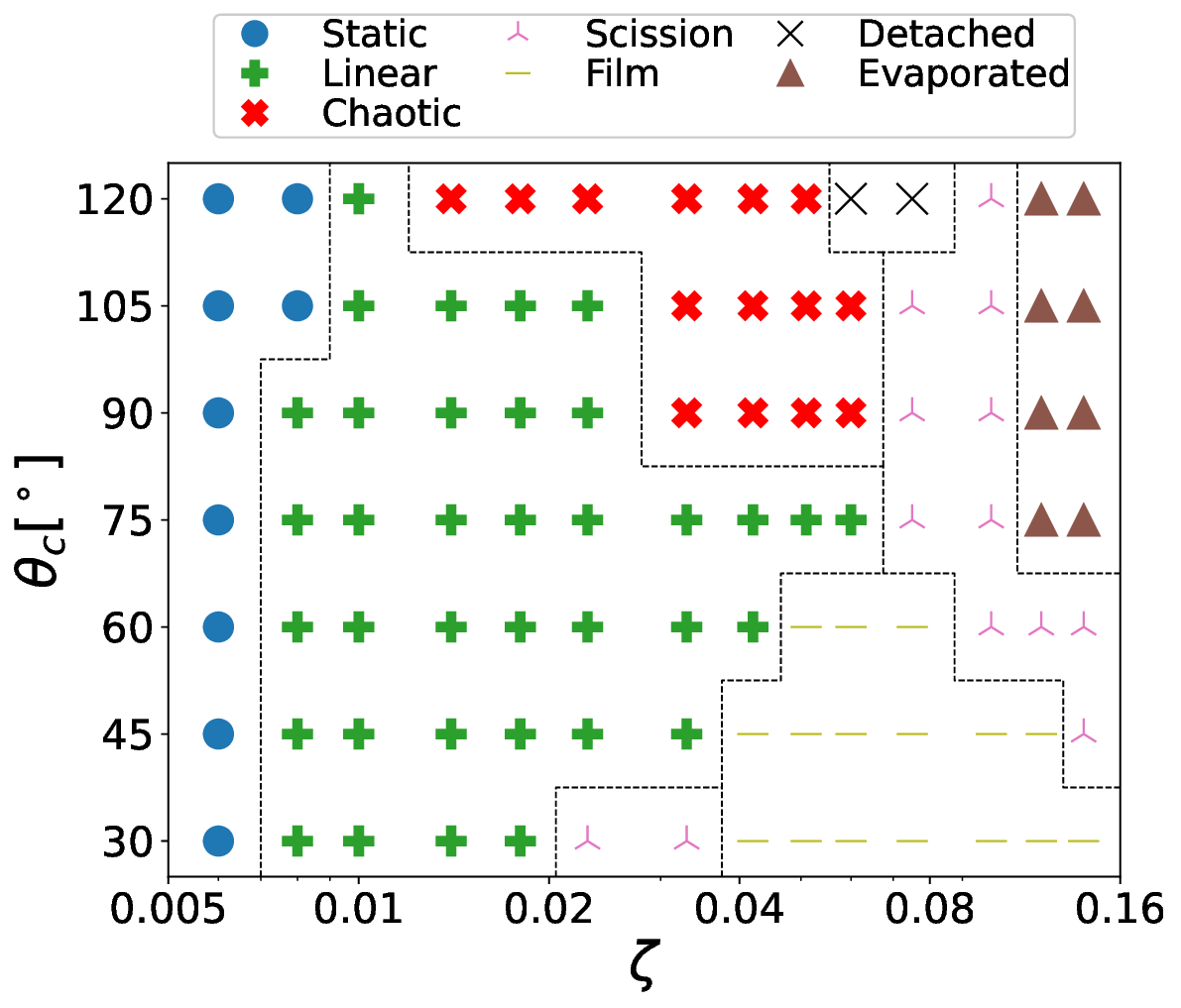}
\caption{Diagram of the droplet dynamical regimes as a function of the activity, $\zeta$, and contact angle, \(\theta_c\), for a homeotropic surface.}
\label{fig:Regimes_Sizes}
\end{figure}

At fixed radius, the droplet dynamical behaviour depends both on the activity and the equilibrium contact angle, which will be explored in detail for homeotropic anchoring. We found distinct dynamical regimes as described below: static, linear, chaotic, scission, spreading, detached and evaporated droplets. These states are identified by analysing the droplet position in time as described in the SM. Figure~\ref{fig:Regimes_Sizes} summarizes the dynamical regimes as a function of the equilibrium contact angle and the activity. We will start by describing the role of activity for a neutral equilibrium contact angle,  $\theta_c=90^\circ$. 
The droplet radius is $R=22.4$.

\begin{figure*}
\centering
\includegraphics[width = 0.9\linewidth]{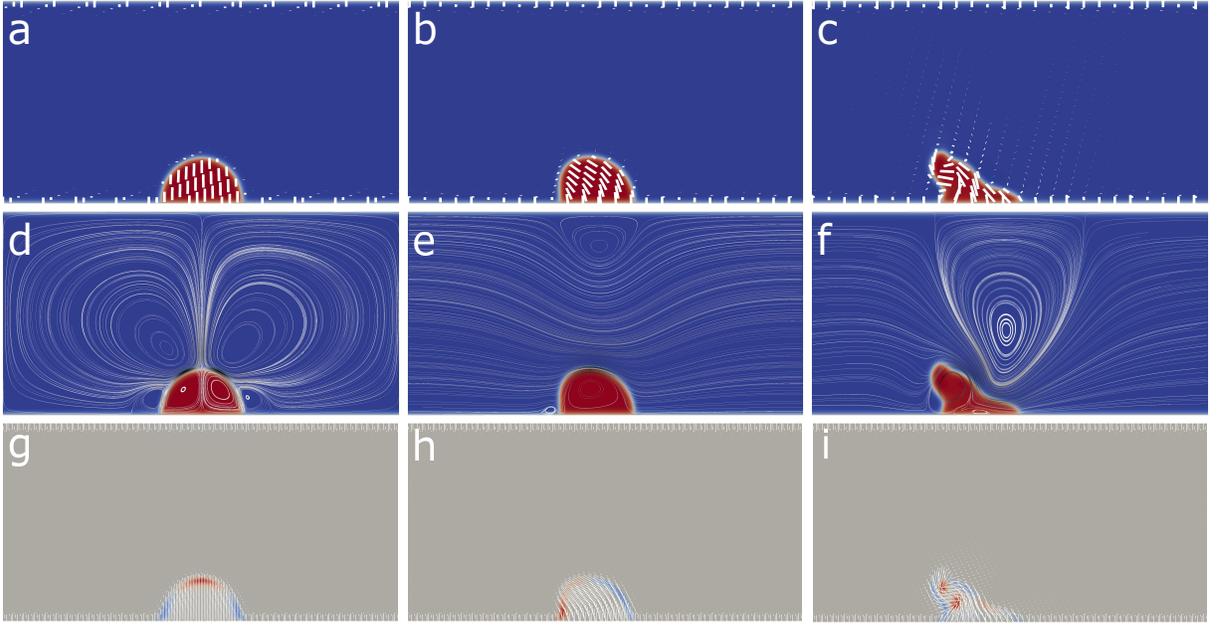}
\caption{Illustration of the director and flow fields for homeotropic surface anchoring with neutral equilibrium contact angle, \(\theta_c = 90^\circ\): static ((\textbf{a}), (\textbf{d}) and (\textbf{g}), \(\zeta = 0.001\)); linear ((\textbf{b}), (\textbf{e}) and (\textbf{h}), \(\zeta = 0.01\)); and chaotic ((\textbf{c}), (\textbf{f}) and (\textbf{i}), \(\zeta = 0.051\)). The top row depicts the director field, the middle row the flow field and the bottom row the charge density field. In (\textbf{d}), the flow velocity has a maximum value of \(1.9 \times 10^{-4}\). In (\textbf{e}), the flow velocity has a maximum value of \(1.5 \times 10^{-3}\). In (\textbf{f}), the flow velocity has a maximum value of \(9.1 \times 10^{-3}\). The streamlines are sketched in the middle panels and the magnitude of the velocity is color coded: white for low and black for high velocities. In the bottom row, the red color represents positive charge density while blue represents negative one. The white lines stand for the director field.}
\label{fig:directVert_examples}
\end{figure*}

\textit{Static} -- The first dynamical regime is characterized by static droplets, and occurs at low activities ($0<\zeta<0.007$) for a neutral contact angle, $\theta_c=90^\circ$. In this state, the droplets remain in their original position with active flows generated close to the droplet (see Fig.~\ref{fig:directVert_examples}(a) and (d)). As a consequence of the flows, which are directed to the droplet on the sides and away from it at the top, the aspect ratio of the droplet changes. A similar behaviour is observed in suspended droplets (far from a surface)~\cite{PhysRevLett.112.147802, PhysRevX.11.021001, Coelho2021ptrsa}. Figure~\ref{fig:directVert_Aratio} reveals that the aspect ratio increases almost linearly with the activity for any equilibrium contact angle. Of course the latter has a weak effect on the slope and on the intercept. In Ref.~\cite{PhysRevLett.112.147802}, the droplet elongation results from the repulsion between two defects at the interface, which are formed by imposing strong homeotropic interfacial anchoring. Although we do not impose any interfacial anchoring, active anchoring arises driven by the active flows. The director field is slightly inclined, with left-right symmetry, as the active anchoring tends to align it parallel to the interface. For droplets in the static regime, this anchoring is weak and a non-singular distortion with positive charge density is formed at the top of the droplet (see Fig.~\ref{fig:directVert_examples}(g)). 

Recall that the passive droplet has an elongation that depends only on the equilibrium contact angle. In what follows, we describe the elongation of the droplet driven by active forces. The director field is now vertical and thus the forces on the sides and on the top of the droplet are given by: 
\begin{align}
\mathbf{F}^s_\perp = -\frac{\zeta \vert \nabla S \vert}{3}\mathbf{m} \quad \text{and} \quad \mathbf{F}^t_\perp = \frac{2\zeta \vert \nabla S \vert}{3}\mathbf{m}.
\label{active-force-homeotropic-eq}
\end{align}
As a result, the droplet is compressed on both sides and stretched in the vertical direction. Notice that this force is per volume and it will act mostly at the interface where the $\vert \nabla S \vert$ is non-zero. Thus, the total force will depend on the droplet radius. As the droplets become more elongated, in the direction perpendicular to the surface, a bend instability occurs at an activity that depends on the contact angle (and on the droplet size, not shown) and the droplet is set in motion. This is why the dotted curves in Fig.~\ref{fig:Regimes_Sizes} end at different activities and shows the dependence of the dynamical regimes on the equilibrium contact angle. 

\textit{Linear} -- At intermediate activities ($0.007<\zeta<0.0155$) and a neutral contact angle $\theta_c=90^\circ$, the symmetry of the homeotropic director field in the droplet, elongated in the direction perpendicular to the surface, is broken and the droplet starts to move with constant velocity, to the right or to the left. Note that, for a neutral contact angle, $\theta_c=90^\circ$, the transition between the static and the linear regimes occurs at $\zeta=0.007$, when $\ell_A\approx 2.4$. The vortex size $\sim 10\ell_A$, is now comparable to the droplet radius, and one vortex fits in the droplet, as illustrated in Fig.~\ref{fig:directVert_examples}(e). We call this the linear regime as the droplet position evolves linearly with time. As shown in Fig.~\ref{fig:directVert_Aratio}, the droplet velocity increases linearly with the activity except at the largest activity where the linear regime starts transitioning to different dynamical regime. The velocity does not change significantly with the contact angle, being slightly larger for smaller contact angles. This dynamical regime is similar to that observed for a droplet on a surface with oblique anchoring (see the Appendix). Although the director field is homeotropic at the surface, this symmetry is broken and the director field becomes oblique elsewhere leading to directed motion, see Fig.~\ref{fig:directVert_examples}(b) and (e)). The symmetric positive charge density in the static regime breaks the left-right symmetry (see Fig.~\ref{fig:directVert_examples}(h)) setting the direction of the self-propelled droplet motion.

\textit{Chaotic} -- At higher activities ($0.0155< \zeta< 0.0675$) and a neutral contact angle, $\theta_c=90^\circ$, the droplet moves randomly and its shape changes in time, see Fig.~\ref{fig:directVert_examples}(c) and (f)). This happens due to the nucleation of pairs of motile defects, within the droplet, which drive spontaneous chaotic flows (see Fig.~\ref{fig:directVert_examples}(i)). As the director field changes randomly in time, the spontaneous flows are also random as found in active turbulence. The droplet is attached to the surface, and thus it can only move in the horizontal direction. Droplets in this state may exhibit directed motion if driven by chemotaxis~\cite{PhysRevE.102.020601} or by asymmetries in the surface~\cite{doi:10.1126/science.aal1979}. On flat uniform surfaces, however, the self-generated flows are random and thus can not drive directed motion.

\begin{figure}
\centering
\includegraphics[width = \linewidth]{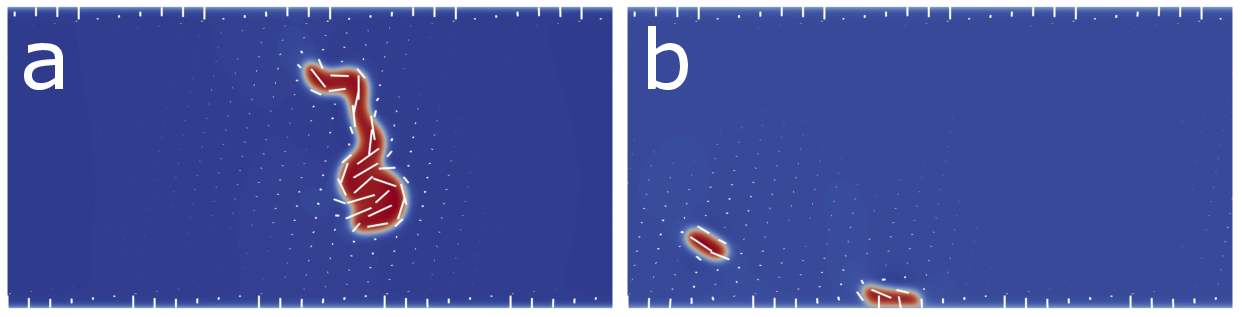}
\caption{Snapshots of two regimes for homeotropic anchoring with non-wetting contact angle \(\theta_c = 120^\circ\): detached ((\textbf{a}), \(\zeta = 0.075\), at \(t = 40000\)); and scission ((\textbf{b}), \(\zeta = 0.1\), at \(t = 960000\)). Although the nematic region is smaller in (\textbf{b}), its total mass is conserved as mixing occurs through local shearing by the active stresses.}
\label{fig:directVert_Broken_Detached}
\end{figure}

\textit{Scission} -- At a threshold activity which depends on the contact angle, ($\zeta_{div}\approx 0.0675$ at a neutral contact angle $\theta_c=90^\circ$), the droplets split and one droplet moves away from the surface. This droplet may attach to one of the two surfaces later, see Fig.~\ref{fig:directVert_Broken_Detached}(b)). This resembles the droplet division regime  reported for suspended droplets, which were shown to divide above a certain activity~\cite{PhysRevLett.112.147802, PhysRevX.11.021001}. Droplet scission occurs as the result of morphological changes (such as fingering and protrusions) driven by the activity rather than by curvature.

\begin{figure*}
\centering
\includegraphics[width = 0.9\linewidth]{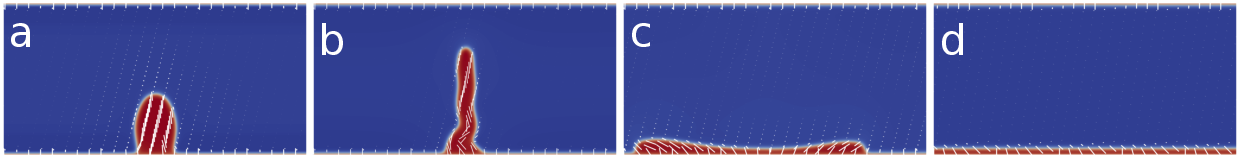}
\caption{Spreading of a droplet with \(\zeta = 0.051\), contact angle \(\theta_c = 45^\circ\) and homeotropic surface anchoring. (\textbf{a}) Initially, the droplet elongates vertically. (\textbf{b}) The elongated droplet is unstable to bend distortions and as a result the director field undergoes strong distortions and takes an oblique random orientation at the surface. (\textbf{c}) The droplet then spreads, as the surface promotes partial wetting (contact angle $\theta_c< 90º$), driven by the active forces. (\textbf{d}) A stable flat film is formed.}
\label{fig:wetLayer_examples}
\end{figure*}

\textit{Film} -- At lower contact angles ($\theta_c<60^\circ$) in the partial wetting regime and high activities ($Ca>10$), the droplet spreads on the surface forming a film-like structure in the steady state. Figure~\ref{fig:wetLayer_examples} illustrates this regime. Initially, the droplet elongates in the direction perpendicular to the surface (driven by the positive defect on the top) but then the director field close to the surface rearranges and becomes oblique, in random directions, and the droplet spreads on the surface. This film-like structure is neither flat nor steady. In the steady state, a wetting nematic film covers the whole surface. The threshold for this flat fim that wraps around the horizontal direction depends on the size of the simulation box and thus we cannot define and active wetting transition. We stress that the apparent contact angle is never zero until finite size effects set in and the film spreads over the whole surface. These effects may be quite drastic. For example, for a surface which is twice the size, the neck becomes so thin that the film breaks and two droplets move away in opposite directions. 

We note that not only the thresholds but the sequence of the dynamical regimes of extensile nematics droplets on flat homeotropic surfaces depend on the equilibrium contact angle. In particular, film-like structures are found in the partial wetting regime while detachment and evaporation occur in the non-wetting regime. 

\textit{Detached} -- In some cases ($\theta_c=120^\circ$, $0.0555 < \zeta < 0.0875 $), in the non-wetting regime, $\theta_c > 90º$, the droplet completely detaches from the surface due to the currents generated by the activity as illustrated in Fig.~\ref{fig:directVert_Broken_Detached}(a). After detaching, the droplet behaves as a chaotic suspended droplet.

\textit{Evaporated} -- At very high activities ($\zeta>0.11$, and a neutral contact angle,  $\theta_c=90^\circ$), the droplet evaporates in the steady state as the local shearing by the active forces overcomes the interfacial tension and the passive and active fluids mix. This happens when the value of the isotropic concentration $\phi_I$, which increases with the activity, reaches the average value of $\phi$ as shown in Fig.~\ref{fig:phiAverage}. This effect was described in Ref.~\cite{PhysRevLett.129.268002} for a similar model as a consequence of active shearing, an effect similar to an increase in temperature or to mechanical stirring of emulsions. The variation of $\phi$ is larger here than for the planar anchoring (Fig.~\ref{fig:horizontal4}(f)) because the droplet is motile and thus the active shearing is stronger. 

\subsection{Droplet size}
\label{dropsize-sec}

\begin{figure}
\centering
\includegraphics[width = 0.9\linewidth]{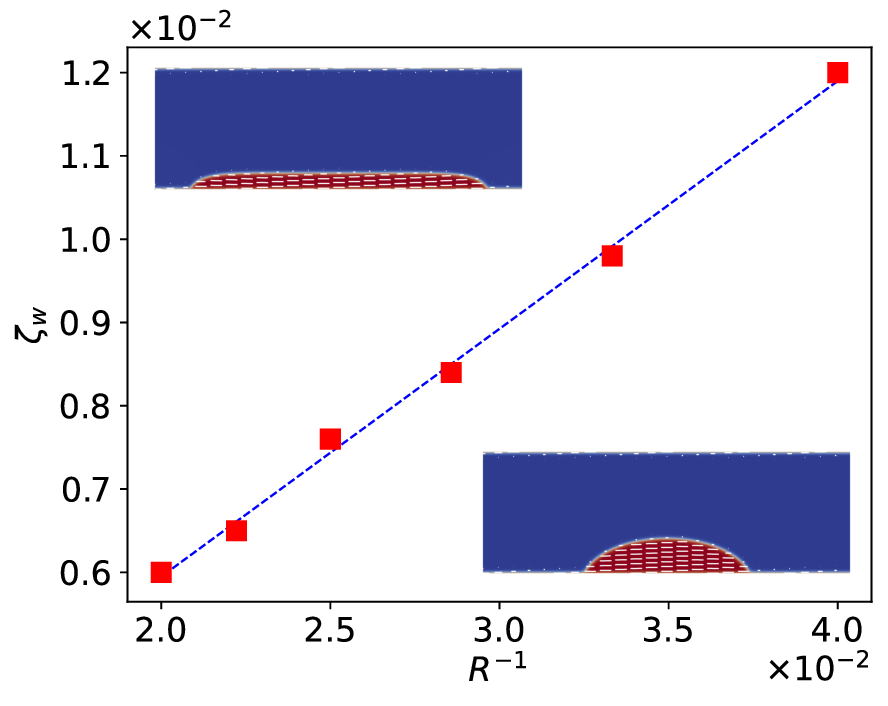}
\caption{Threshold activity for the active wetting transition of a droplet on a surface with planar anchoring and partial wetting equilibrium contact angle $\theta_c=60^\circ$. The dashed line is a linear fit $\zeta_{w}(R^{-1})= a R^{-1}$, where $a=0.30$. We considered radii from $R=25$ to $50$. The insets illustrate the typical droplet shape above and below the transition line. The inset on the top is for a droplet with radius $R=40$ and $\zeta=0.0106$ and that on the bottom is for a droplet with the same radius and $\zeta=0.0028$.}
\label{fig:wetsize}
\end{figure}

\begin{figure}
\centering
\includegraphics[width = 0.9\linewidth]{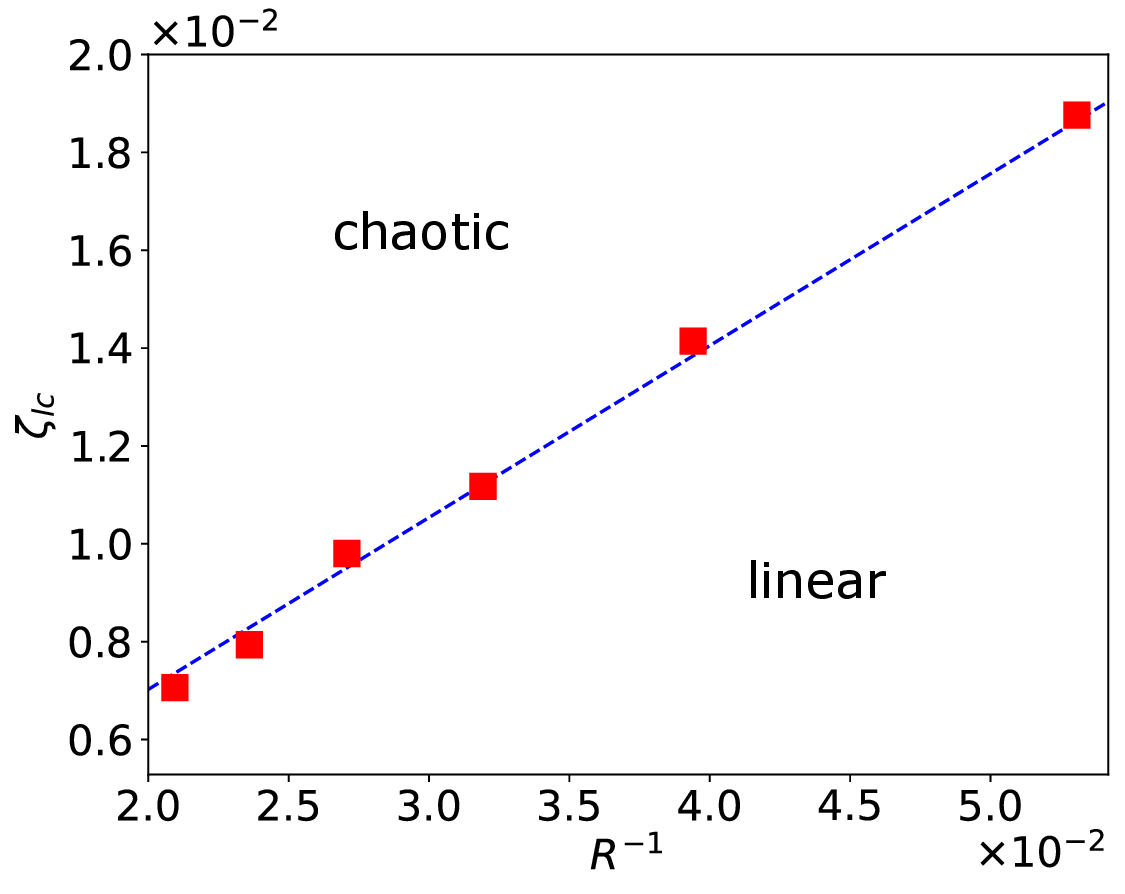}
\caption{Threshold activity for the transition from linear to chaotic motion of a droplet on a homeotropic surface with neutral equilibrium contact angle, $\theta_c=90^\circ$, as a function of the inverse radius, $R^{-1}$. The dashed line is a linear fit: $\zeta_{lc}(R^{-1})= a R^{-1}$, with $a=0.35$. We considered radii from $R=25$ to $50$. }
\label{fig:directVert_zR_linear}
\end{figure}

The shape of passive nematic droplets may depend weakly on their size and the elastic anisotropy. If the radius of the droplet is much larger than the nematic correlation length, $\ell_N$, and there is interfacial anchoring due to elastic anisotropy, the droplet is slightly elongated in the direction parallel to the director field. In the simulations reported here, this effect is absent as we neglected elastic anisotropy (single elastic constant approximation). Thus, the simulated passive droplets are always circular.

It has been suggested that the threshold of the active wetting transition depends on the droplet size~\cite{PrezGonzlez2018, doi:10.1126/science.abo5423} and the estimated active forces discussed earlier scale with $R \zeta$. We stress that the arguments leading to this scaling assume that the droplet is away from the active turbulent and other non-steady regimes. 

We start with finite droplets and  investigate the size dependence of the dynamical transitions on the droplet size. We analyse the thresholds of two distinct transitions: partial to complete wetting on a planar surface and linear to chaotic motion on a homeotropic surface. 

In Fig.~\ref{fig:wetsize} we plot the threshold activity $\zeta_w$ where the apparent contact angle vanishes for droplets of different radii. This is determined by the minimum activity where at least $60\%$ of the droplet's perimeter is flat (see the insets of Fig.~\ref{fig:wetsize}).  The linear fit supports the assumption that $\zeta R$ is constant at the active wetting transition (on a given surface) and, the slope yields the active wetting threshold capillary number for extensile active nematics on a surface with strong planar anchoring and a partial wetting equilibrium contact angle, $\theta_c=60^\circ$: $Ca_\alpha^{w} \approx 5$. This linear fit leads to $\zeta_w\approx 0.013$ for a droplet of radius $R=22.4$, which is close to the activity where the curve in Fig.~\ref{fig:horizontal4}(e) deviates from the linear regime.

Likewise, Fig.~\ref{fig:directVert_zR_linear} depicts the threshold activity $\zeta_{lc}$ at the transition from linear to chaotic motion, for droplets with different radii. The surface is homeotropic and the equilibrium contact angle is neutral. The data follows a linear relation, yielding the linear to chaotic motion threshold capillary number for extensile active nematics on surfaces with strong homeotropic anchoring and neutral equilibrium contact angle, $\theta_c=90^\circ$: $Ca_\alpha^{lc} \approx 5.8$.

In both cases, the linear relation between the threshold activity and the inverse radius supports the assumption that the active capillary number $Ca_\alpha$ controls the dynamics of active nematics droplets on a flat surface at a fixed $\theta_c$. Of course, surface effects such as the surface anchoring and the equilibrium contact angle are not encoded in the active capillary number, $Ca_\alpha$, and different capillary numbers at these thresholds will be found for different surfaces. It is still remarkable that the dynamics of droplets with different activities and radii, on a particular surface, are controlled by a single physical parameter. 

This simple picture, however, will change for sufficiently large droplets where active turbulence sets in at vanishingly small activities. This has drastic consequences for the active wetting and other dynamical transitions, as we will discuss in the conclusion.

\section{Conclusion}

We considered the dynamics of extensile active nematics droplets on flat surfaces, based on the continuum hydrodynamic theory. 
We investigated a range of dynamical regimes as a function of the surface anchoring, equilibrium contact angle, activity and droplet radius. 

The first two parameters (surface anchoring and equilibrium contact angle) characterize the surface-fluid interactions and we have considered surfaces with planar, homeotropic and oblique (in the Appendix) anchoring in the strong regime, as well as zero anchoring (in the Appendix). The solid-fluid interactions are further characterized by the equilibrium contact angle, $\theta_c$.  

We have shown that the total NI interfacial tension, $\Sigma$, and thus the equilibrium contact angle is dominated by the spatial variation of the concentration profile across the interface, as the contribution of the orientational order parameter profile is less than $0.5\%$. This can be understood by recalling that the NI transition is weakly first-order while the concentration field is considered to be deep in the bulk two-phase region. 

When the activity is switched on, we found that for finite droplets there is a single physical parameter $\zeta R/\Sigma$ that controls the dynamics on a specific flat surface, in line with previous results that revealed that this parameter, coined as the active capillary number, $Ca_\alpha$, controls the dynamical behaviour of suspended droplets.    

We found a wide range of dynamical regimes of active droplets on flat surfaces, including static droplets, self-propelled linear motion, chaotic motion, droplet scission, active wetting and droplet detachment and evaporation. The thresholds and sequence of these dynamical regimes depend of the surface anchoring and on the equilibrium contact angle.

The last three regimes were not reported for suspended droplets as wetting and detachment require the presence of a surface and droplet evaporation requires an activity driven shift of the coexisting concentrations, which was hindered or suppressed  in Ref.~\cite{PhysRevLett.112.147802} by imposing volume (or area) conservation of the nematic phase. 

We found that the nematic order parameter in active droplets varies with the activity, affecting the wetting behaviour in a way that resembles the original argument by John Cahn for critical point wetting. In the present context it turns out that this effect is sub-dominant but this is ultimately responsible for the observed evaporation of the droplets. 
The dominant mechanism driving active wetting is related to the generation of active forces parallel to the surface that result from gradients in the nematic order parameter. These forces are proportional to the droplet radius $R$ and point outwards on both sides of extensile active droplets. On planar surfaces there is a well defined threshold where the apparent contact angle vanishes. On homeotropic surfaces, by contrast, the spreading film undulates and it becomes flat only when it covers the whole surface, as a result of the finite size of the simulation domain. 

Fingering instabilities prior to an active wetting transition were reported in~\cite{PrezGonzlez2018} and may be characteristic of transient states as those described above for homeotropic surfaces before active spreading sets in. Note that the presence of shape instabilities does not appear to be a necessary condition for active wetting, as it was not observed for agonistic surface and active anchorings (planar surface and active anchorings) but may play a role when these anchorings are antagonistic (homeotropic surface and planar surface anchorings). The droplet size will also affect the onset of these shape instabilities, as for large droplets the vortices characteristic of active turbulence are no longer screened by the confinement of the active nematic.  

In the simplest case (planar-planar) we have estimated the active wetting threshold by the vanishing of the apparent contact angle and checked that this was independent of the size of the simulation domain. Strictly speaking, however, in the infinite size limit the active nematic is turbulent at any activity and it does not coexist with an isotropic phase. In that limit, the arguments discussed above as well as those proposed in published work to interpret experimental observations of active wetting do not apply. The active wetting transition is a finite size dynamical transition and it cannot occur in the infinite droplet size limit. Even the droplet size effect captured by the active wetting threshold dependence on the active capillary length suggests that for infinite droplets the threshold activity at the wetting transition vanishes.  

The idea of an active wetting transition, however, is a useful concept that has already been shown to describe experimental observations both in wet as well as in dry active systems~\cite{doi:10.1126/science.abo5423} and~\cite{PrezGonzlez2018}.

The experimental system considered in~\cite{doi:10.1126/science.abo5423} is well described by our model of an extensile active-passive mixture on a flat surface with planar anchoring. In fact, the theoretical model used in~\cite{doi:10.1126/science.abo5423} 
is identical to ours except that it does not impose mass conservation. That allows the growth of wetting films of any thickness but 
it is not clear how that thickness is set in a self-consistent manner. Having noted this difference, the film thickness at the capillary wall in~\cite{doi:10.1126/science.abo5423} plays the role of the droplet radius in our model. Other effects related to mass conservation, such as the droplet evaporation will not be described by the model used in~\cite{doi:10.1126/science.abo5423}.

The connection with the experiments and theory that revealed an active wetting transition between 2D epithelial monolayers and 
3D aggregates~\cite{PrezGonzlez2018} is not as straightforward. To begin with the system is dry so that hydrodynamics does not play a role. The model is also quite different as there is no underlying thermodynamic transition at zero activity. The model considers, by contrast, an interplay between two different active forces the ratio of which defines an intrinsic lengthscale that controls active wetting. In more recent work the model was further elaborated~\cite{Pallars2022} but these essential differences remain. 

In summary, our analysis provides an overview of the striking dynamical regimes reported for active droplets on surfaces, ranging from self-propulsion to droplet evaporation and gives a unified description of active wetting of finite active droplets, in systems with only one active force or where the effect of one of the active forces dominates.  

Although active wetting is ultimately a finite droplet size dynamical transition, it may be well characterized and was proved useful in the interpretation of experimental results. 

\section*{Conflicts of interest}
There are no conflicts to declare.

\section*{Acknowledgements}
We acknowledge financial support from the Portuguese Foundation for Science and Technology (FCT) under the contracts: EXPL/FIS-MAC/0406/2021, PTDC/FISMAC/5689/2020, UIDB/00618/2020 and UIDP/00618/2020. We thank J. M. Romero-Enrique for the fruitful discussion.

\section*{Appendix}

\subsection*{Landau-de Gennes Interface}

\subsubsection*{Interfacial profiles}

We consider a flat NI interface with homeotropic anchoring. Under these conditions, both the interface and the bulk nematic are uniaxial and only the scalar orientational order parameter $S$ varies across the interfacial region, which has an intrinsic width of the order of the nematic bulk correlation length, $\ell_N$~\cite{p1995physics}. The Landau-de Gennes free energy density, Eq.~\eqref{fldg-eq}, is written now in powers of $S$ and its derivative in the direction perpendicular to the interface, say $x$:
\begin{align*}
 f_{LdG} &= \frac{A_0}{2}\left( 1- \frac{\gamma}{3} \right) Q_{\alpha \beta}^2 - \frac{A_0\gamma}{3} Q_{\alpha \beta} Q_{\beta \gamma} Q_{\gamma \alpha} \\
  & + \frac{A_0\gamma}{4} (Q_{\alpha \beta} Q_{\alpha \beta} )^2 + \frac{L}{2} (\partial _\gamma Q_{\alpha \beta})^2\\
 &= \frac{A_0}{3}\left( 1-\frac{\gamma}{3}\right)S^2 - \frac{2A_0\gamma}{27}S^3 + \frac{A_0\gamma}{9}S^4 + \frac{L}{3}\left( \frac{d S}{d x} \right)^2.
\end{align*}
where the numerical coefficients in the second line arise from the sum over the indices of $Q_{\alpha \beta}$ implicit in the first.

This free energy has two minima at:  
\begin{align}
S_I=0 \quad \text{and} \quad S_N=\frac{\gamma + \sqrt{3(-8\gamma+ 3\gamma^2)}}{4\gamma}.
\label{sn-eq}
\end{align}
one of which is stable and the other metastable depending on the value of the parameter $\gamma$. The value of the free energy density vanishes at both minima when $\gamma=2.7$, where a nematic phase with scalar order parameter $S_N=1/3$ coexists with the isotropic phase with $S_I=0$.

\begin{figure}[htb]
	\includegraphics[width =\linewidth]{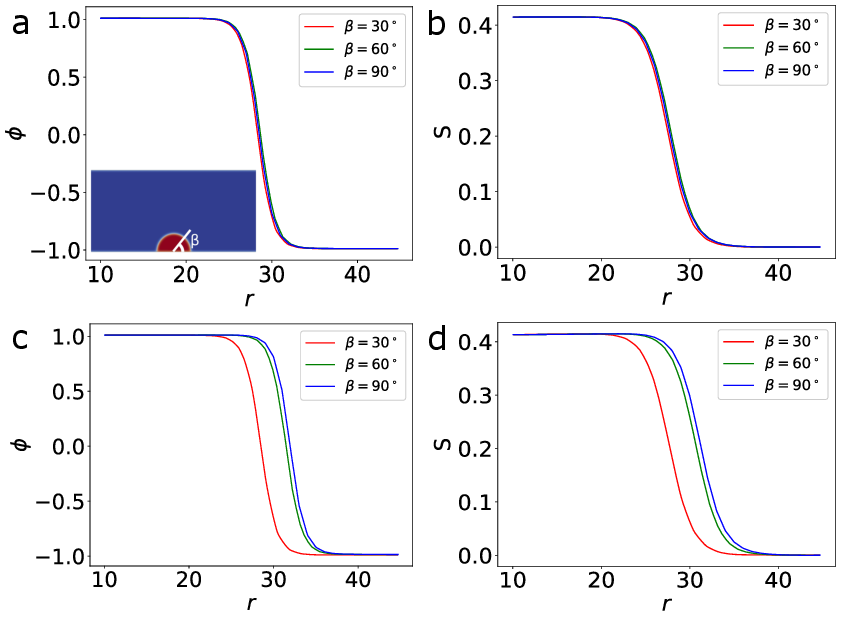}
	\caption{Concentration \(\phi\) and orientational order parameter \(S\) for two static extensile active nematics droplets, with $R=22.4$, on a homeotropic surface with a neutral equilibrium contact angle, \(\theta_c = 90^\circ\). The profiles were measured along the line at an angle \(\beta\) with the surface (see the inset). The radial distance \(r\) is measured from the center of the droplet on the surface. Top row, (\textbf{a} and \textbf{b}), droplet with activity \(\zeta = 0.001\). Bottom row, (\textbf{c} and \textbf{d}), droplet with activity \(\zeta = 0.006\).}
	\label{s-phi-profile-fig}
\end{figure}
\begin{figure}[htb]
	\includegraphics[width = \linewidth]{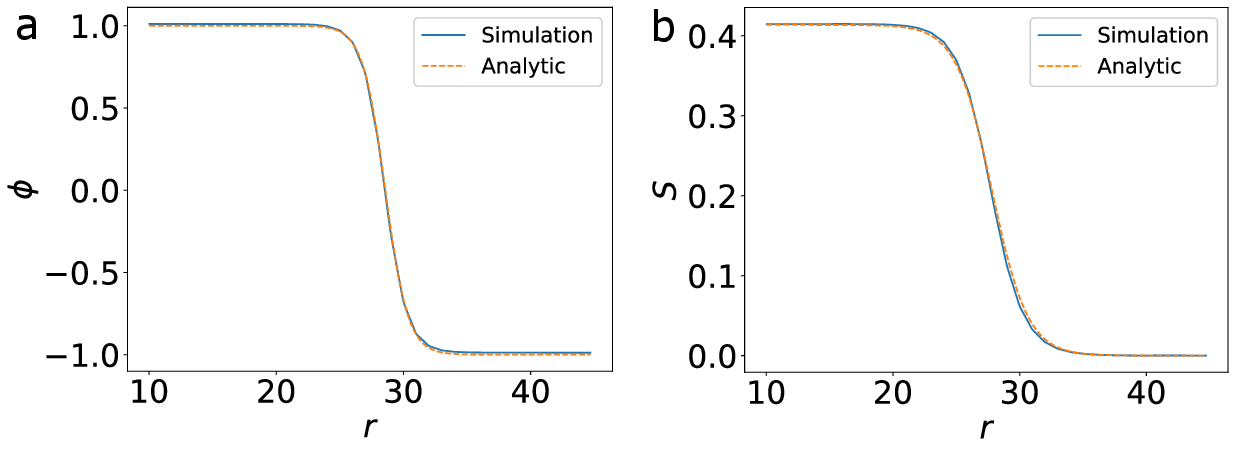}
	\caption{Comparison between the analytic and simulated profiles \(\phi\) (\textbf{a}) and \(S\) (\textbf{b}) along the vertical  \(\beta = 90 ^\circ\) direction, for a droplet with activity \(\zeta = 0.001\) on a homeotropic surface with a neutral equilibrium contact angle, \(\theta_c = 90 ^\circ\).}
	\label{teo-an-fig}
\end{figure}
The equilibrium order parameter profile, obtained by minimizing the total free energy, $\mathcal{F}=\int f d^3 r$, is the solution of the ordinary differential equation:
\begin{align*}
\frac{\delta \mathcal{F}}{\delta{S}}&= \frac{2A_0}{3}\left( 1-\frac{\gamma}{3}\right)S - \frac{2A_0\gamma}{9}S^2 + \frac{4A_0\gamma}{9}S^3 \\
&- \frac{2L}{3}\frac{d^2 S}{d x^2}  =0.
\end{align*}
A closed form solution may be obtained as follows~\cite{PhysRevE.86.011703}. Multiplying the equation for $S(x)$  by $dS/dx$ and integrating from $-\infty$ to $x^\prime$, we find:
\begin{align*}
&\int^{x^\prime}_{-\infty} \frac{2L}{3} \frac{d^2 S}{d x^2} \frac{d S}{d x} dx =\\
& \int^{x^\prime}_{-\infty} \left[ \frac{2A_0}{3}\left( 1-\frac{\gamma}{3}\right)S - \frac{2A_0\gamma}{9}S^2 + \frac{4A_0\gamma}{9}S^3 \right] \frac{d S}{d x} dx\\
& \Leftrightarrow \frac{L}{3}\left( \frac{dS}{dx} \right)^2=\frac{A_0}{3}\left( 1-\frac{\gamma}{3}\right)S^2 - \frac{2A_0\gamma}{27}S^3 + \frac{A_0\gamma}{9}S^4,
\end{align*}
where we assumed that $dS/dx$ vanishes at $-\infty$. The two phases coexist at $\gamma=2.7$, where:
\begin{align}
&\frac{L}{3}\left( \frac{dS}{dx} \right)^2=\frac{A_0}{30}S^2(3S-1)^2 \label{eq1}\\
& \Leftrightarrow \int \frac{1}{S(S-1/3)} dS = -3x\sqrt{ \frac{A_0}{10L}} +C,\nonumber
\end{align}
with the integration constant $C=0$ setting the interface at $x=0$. Finally, the explicit solution is found through the change of variable $u=S-1/6$, 
\begin{align}
&\int \frac{1}{u^2-1/36}du \nonumber\\ 
&= -6\tanh^{-1}(6u)=-6\tanh^{-1}(6S-1) = -\sqrt{\frac{A_0}{10L}}3x \nonumber\\
& \Leftrightarrow S(x) = \frac{S_N}{2}\left[ 1+ \tanh\left( \frac{x}{2\ell_N} \right) \right], \label{s-prof-eq}
\end{align}
where $S_N=1/3$ is the nematic order parameter at coexistence and $\ell_N=\sqrt{\frac{10L}{A_0}}$ is the nematic correlation length under the same conditions.

\begin{figure}[h]
	\includegraphics[width = 0.9\linewidth]{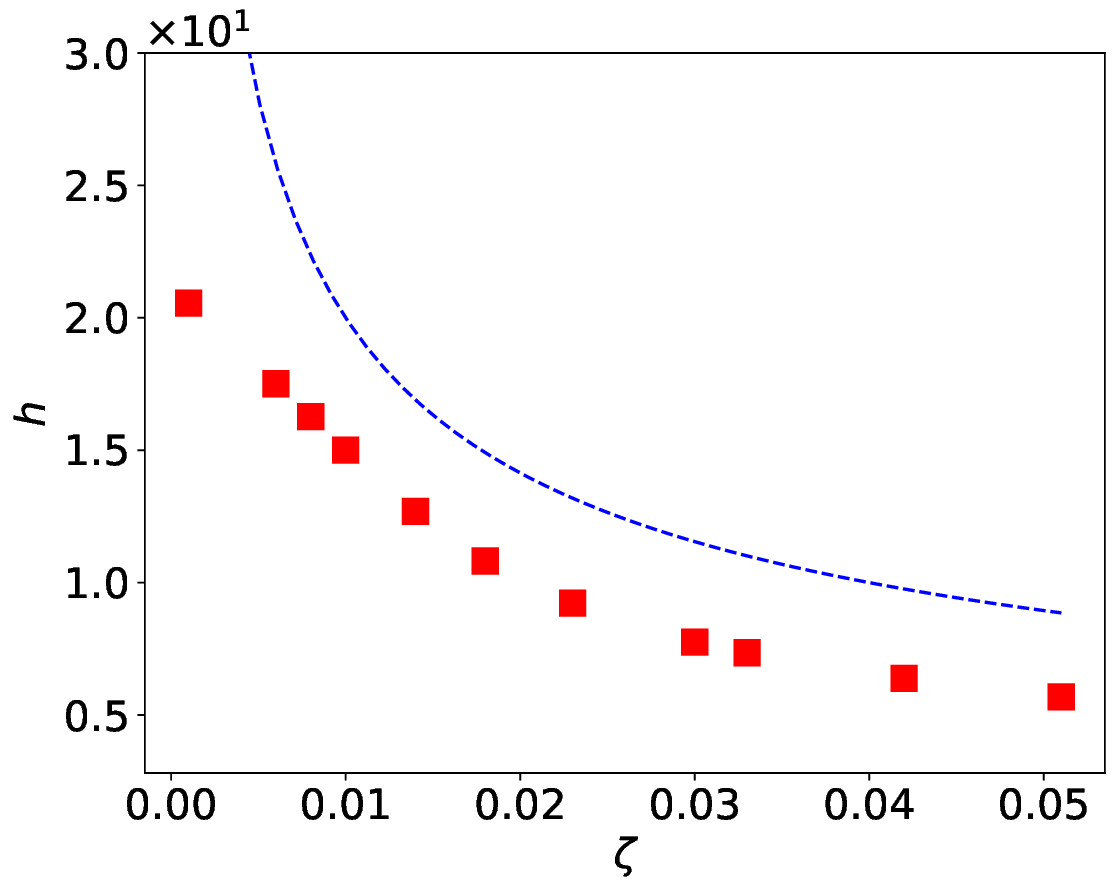}
	\caption{ Height of the droplet as a function of the activity. Droplet on a surface with planar anchoring, with equilibrium partial wetting contact angle, $\theta_c=60^\circ$.  A typical vortex size ($\sim 10\ell_A$) is plotted as a blue dashed line for comparison. }
	\label{planar-fig}
\end{figure}

We proceed to compare the composition and orientational order parameter profiles of large droplets (radii at least a factor of 10 larger than the concentration and nematic correlation lengths) at low activities obtained from the simulations with the analytical results for the decoupled passive concentration and orientational order parameter profiles, Eqs.~\eqref{phi-eq-profile} and Eq.~\eqref{s-prof-eq}. Note that although Eq.~\eqref{s-prof-eq} for the density profile is obtained at coexistence ($\gamma=2.7$) we assume the same form and simply change the value of the nematic order parameter $S_N$ and the nematic correlation length $\ell_N$ at the new value of the parameter $\gamma$ elsewhere in the nematic phase. 
 
We start by analysing the numerical results for $\phi(r)$ and $S(r)$ plotted in Fig.~\ref{s-phi-profile-fig} for extensile active nematics droplets with $R=22.4$ at two different activities $\zeta$, on a homeotropic surface with a neutral equilibrium contact 
angle, $\theta_c=90º$. In the absence of activity the droplets are circular but as the activity increases active anchoring promotes parallel alignment at the interface and the droplet elongates in the vertical direction. We have plotted the profiles along three different angles $\beta$ (measured at the center of the droplet as shown in the inset). For the low activity droplet, the profiles $\phi$ and $S$ hardly change with the angle, as the droplet remains nearly circular. As the activity increases, the droplet elongates and the position of the interface ($S=S_N/2$) changes with $\beta$. However, the profiles are very similar, the most noticeable change being a shift in $r$. 

In Fig.~\ref{teo-an-fig} we compare the theoretical concentration $\phi$ and orientational order parameter $S$ profiles for a passive flat interface with those of a droplet at low activity. Equation~\eqref{sn-eq} gives $S_N$ at $\gamma=2.8$ and we used Eq.~\eqref{s-prof-eq} for the orientational order parameter profile at the value of $S_N$ off coexistence. The correlation lengths are (for $\gamma=2.8$): $\ell_N=2.45$ and $\ell_\phi=1.26$. One can see from Fig.~\ref{teo-an-fig} that for large droplets and low activities, the profiles are almost identical to those of a passive flat interface, where the concentration and the orientational order parameter profiles are decoupled. 

\subsubsection*{Interfacial tension}

At coexistence, the interfacial tension of the NI interface is given by:
\begin{align*}
\Sigma_N &= \int^{\infty}_{-\infty}  f_{LdG}(\gamma=2.7) dx , \quad\text{from Eq.~\eqref{eq1}:}\\
& = \frac{2L}{3}\int^{\infty}_{-\infty}\left(\frac{dS}{dx} \right)^2 dx, \quad\text{using Eq.~\eqref{eq1} again:}\\
&= -\frac{2}{3}\sqrt{\frac{A_0 L}{10}} \left( S^3 - \frac{S^2}{2} \right\vert^{S_N}_0 , \quad \text{but }S_N=1/3\\
&= \frac{\sqrt{A_0 L}}{81\sqrt{10}}.
\end{align*}

The contribution from the $Q_{\alpha\beta}$ to the interfacial tension is near-critical and thus much smaller than the contribution from $\phi$. This is corroborated by inspection of the order-parameter profiles for passive and active droplets in Figs.~\ref{s-phi-profile-fig} and~\ref{teo-an-fig}, where the width of the S profile is $\approx 2.8$ times larger than that of $\phi$. We recall that the coefficient of the quadratic term of the LdG free energy density Eq.~\eqref{fldg-eq} varies around its value at the NI transition in the passive system and this implies that the nematic order parameter field varies on longer lengthscales than the concentration field, which is deep in the phase separated regime. This suggests that we may consider only the contribution of $\Sigma_\phi$ to the surface tension, i.e.,  $\Sigma=\Sigma_N+\Sigma_\phi \approx \Sigma_\phi$.

\subsection*{Planar and homeotropic anchoring}

\begin{figure}[htb]
\centering
\includegraphics[width = 0.9\linewidth]{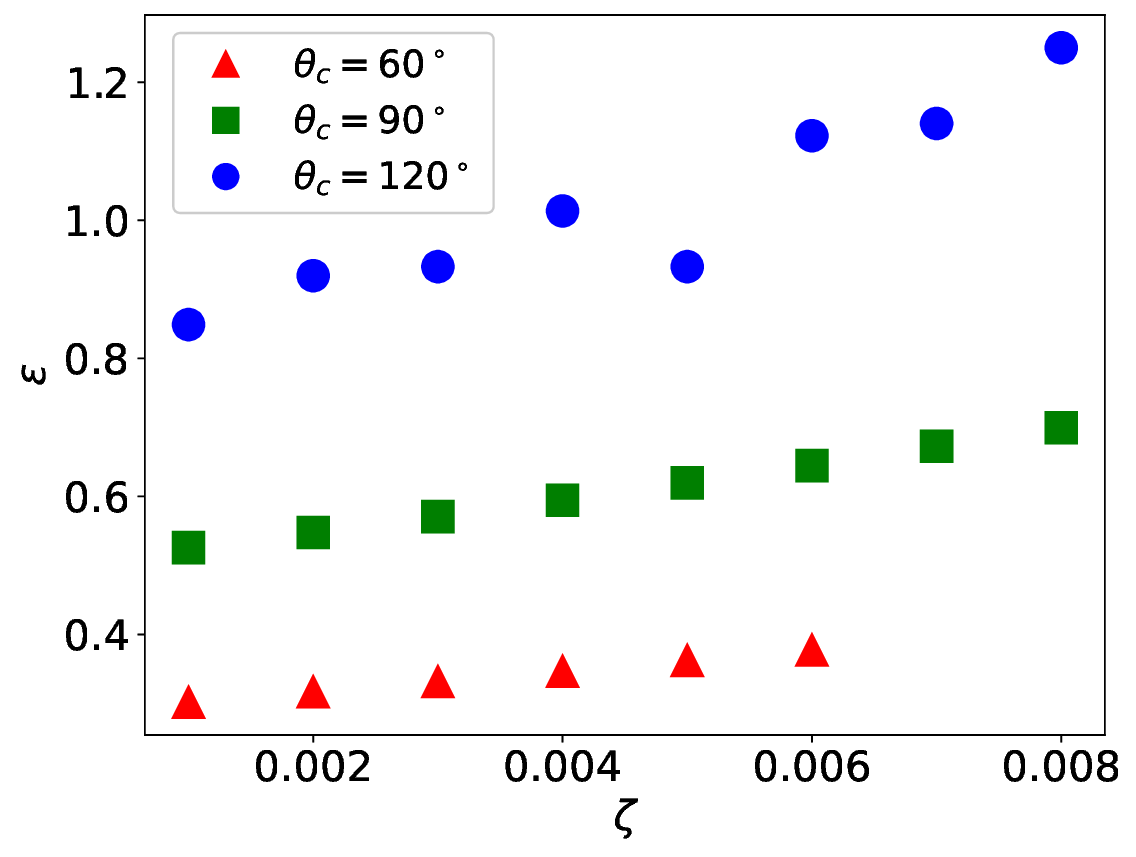}
\includegraphics[width = 0.9\linewidth]{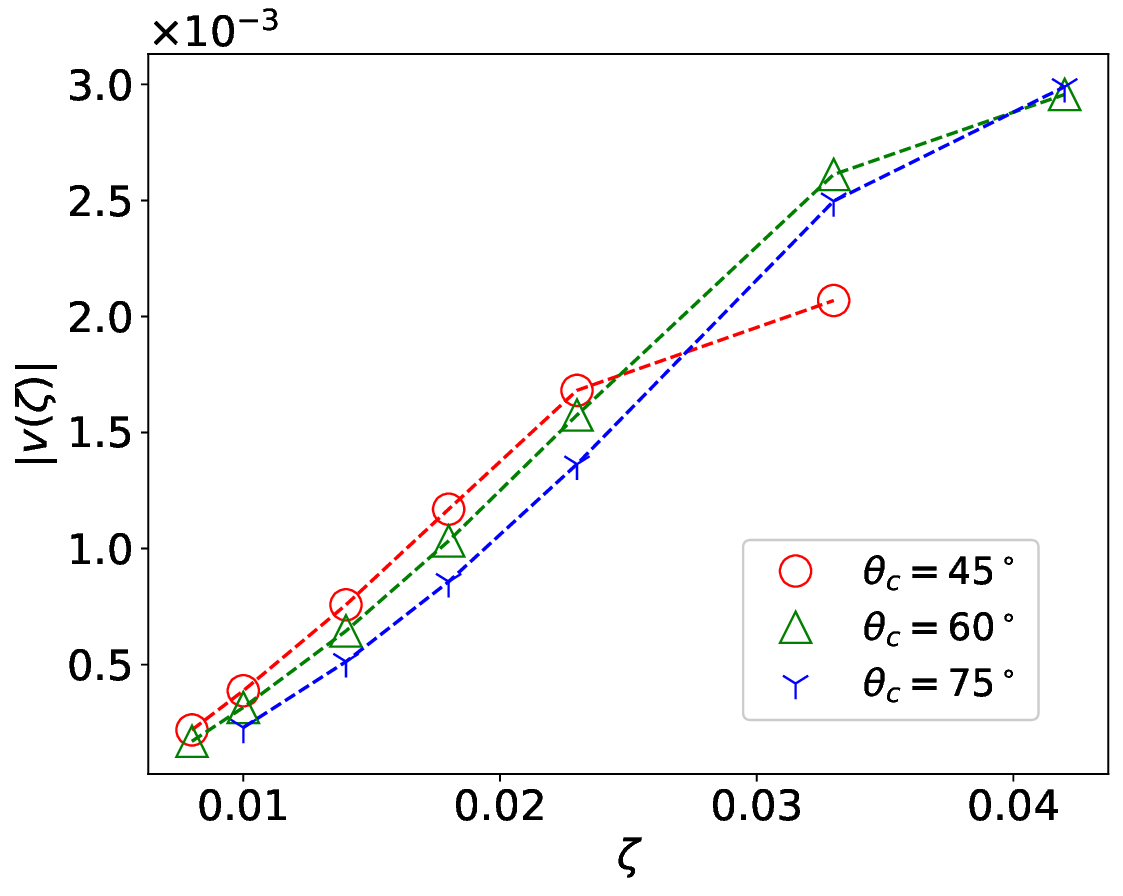}
\caption{(top) Aspect ratio of static droplets as a function of the activity, $\zeta$, on homeotropic surfaces with different contact angles \(\theta_c\). (bottom) Absolute velocity of the droplets in the linear dynamical regime as a function of the activity, $\zeta$, on homeotropic surfaces with different contact angles \(\theta_c\).  }
\label{fig:directVert_Aratio}
\end{figure}

In Fig.~\ref{fig:horizontal4}(e), we plot the aspect ratio of the droplet on a substrate with planar anchoring as a function of the activity. It is also useful to plot the droplet height as shown in Fig.~\ref{fig:directVert_Aratio}. Initially, the height decreases linearly and then the slope changes around $\zeta_{w}\approx 0.015$, as for the aspect ratio. As discussed in the main text, this change can be used to estimate the wetting transition threshold. For $\zeta>\zeta_w$, the top of the droplet becomes flat and the apparent contact angle vanishes. The dashed line in Fig.~\ref{fig:directVert_Aratio} stands for the vortex size $\approx 10\ell_A$. The droplet height is always smaller than the vortex size, which explains why the droplet does not become turbulent: vortices can not form inside the droplet.

For surfaces with homeotropic anchoring, the droplet elongates perpendicular to the surface. Figure~\ref{fig:directVert_Aratio} shows that the aspect ratio (heigh/width) increases with the activity in the static regime. This occurs independently of the contact angle. At higher activities, the droplet transitions to the linear regime, characterized by motion at constant velocity and shape. Figure~\ref{fig:directVert_Aratio} shows the droplet velocity as a function of the activity. The velocity increases with the activity almost linearly except at the end of the curves (at high activities) where the transition to the chaotic regime occurs. 

\begin{figure}
\centering
\includegraphics[width = \linewidth]{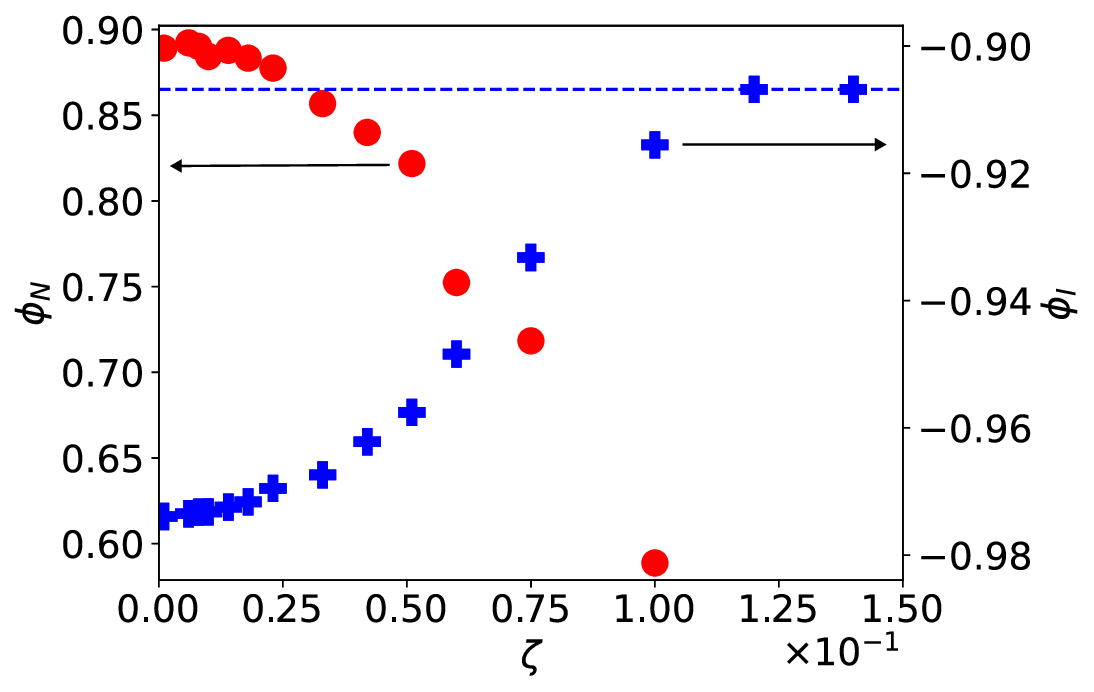}
\caption{Average value of $\phi$ for droplets with homeotropic anchoring and $\theta_c=90^\circ$ as a function of the activity, $\zeta$, in the nematic phase ($\phi_N$, on the left, red circles) and in the isotropic phase ($\phi_I$, on the right, blue crosses). The coexisting values of $\phi$ at activities close to zero are slightly different from $\pm 1$ due to the non-zero thickness of the interface. The dashed blue line represents the average value of $\phi$ (axis on the right), which is constant.}
\label{fig:phiAverage}
\end{figure}

The active stress promotes mixing of the two components (isotropic and nematic fluids). Figure~\ref{fig:phiAverage} depicts the change in the mean values of $\phi$ for the nematic ($\phi_N$) and isotropic ($\phi_I$) components as a function of the activity for a droplet on a homeotropic surface. The concentrations $\phi_N$ and $\phi_I$ deviate from their equilibrium values in the passive mixture $\phi_0=\pm 1$ and become closer as the activity increases. At higher activities, the droplet evaporates and the concentration becomes uniform throughout the system, which remains in a single isotropic phase. This happens when the value of $\phi_I$, which increases with the activity, reaches the average value of $\phi$ as shown in Fig.~\ref{fig:phiAverage}. The variation of $\phi$ with the activity is larger for homeotropic than for planar anchoring (Fig.~\ref{fig:horizontal4}(f)) as the velocities are higher and thus mixing is more effective. In particular, in the chaotic state shearing is much stronger than in the static regime with planar anchoring.


\subsection*{Oblique anchoring}
\label{sec:sec:directIncl}

\begin{figure}[htb]
\centering
\includegraphics[width = 0.9\linewidth]{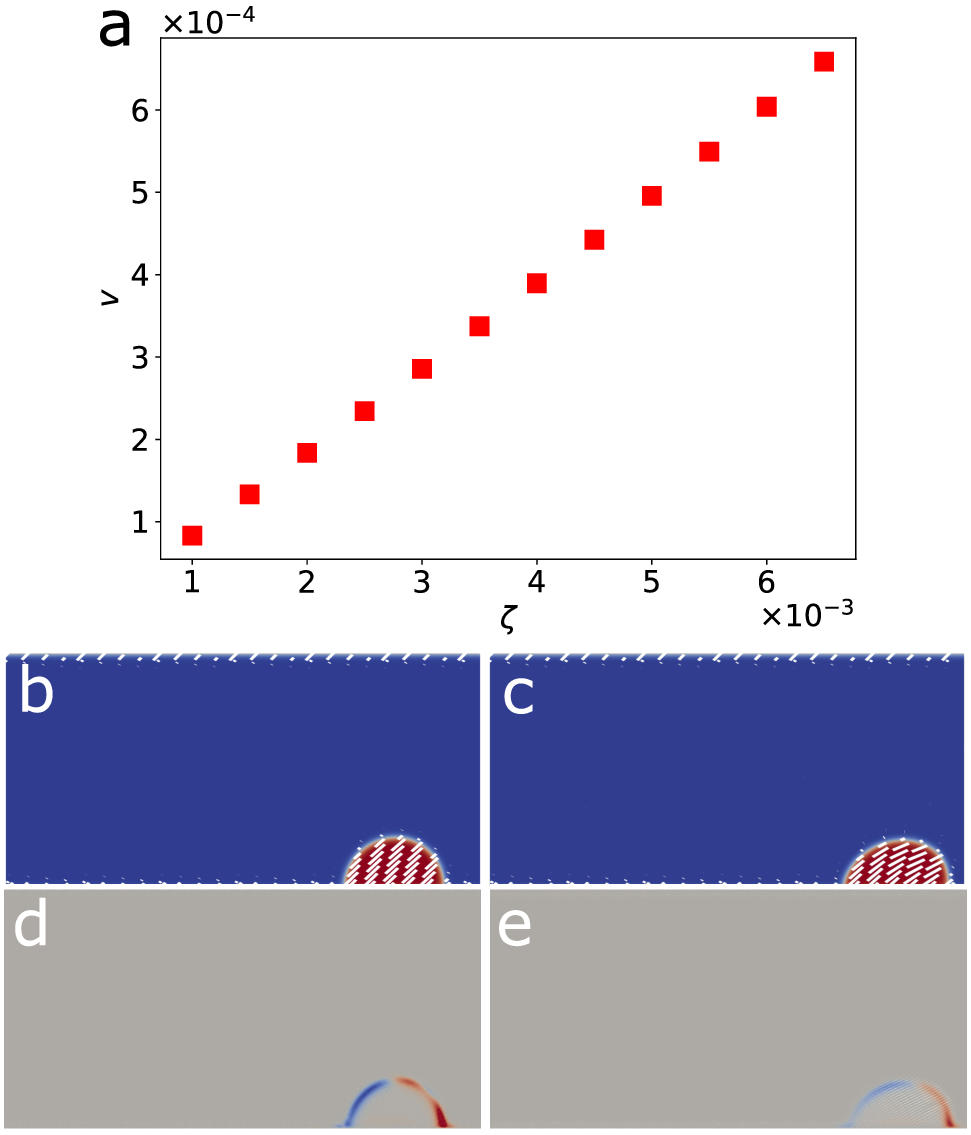}
\caption{(\textbf{a}) Absolute value of the droplet velocity as a function of the activity, $\zeta$, for oblique surface anchoring and a neutral equilibrium contact angle, \(\theta_c = 90^\circ\). (\textbf{b}) and (\textbf{c}) Snapshots of the droplets at the extremes of the curve in (a) ($\zeta=0.001$ and $\zeta=0.0065$ respectively), which are moving to the right with constant velocity and shape. (\textbf{d}) and (\textbf{e}) Charge density in the two droplets shown in (b) and (c). Red represents a positive charge density while blue represents a negative one.}
\label{fig:directIncl_Velocity}
\end{figure}

When the surface anchoring is oblique (at $45^\circ$), we find that the droplet moves with constant velocity at any activity. This happens as the left-right symmetry of the nematic director field in the droplet is explicitly broken by the surface anchoring and as a result the velocity field generated by the activity drives the droplet motion. In Fig.~\ref{fig:directIncl_Velocity}(a), we plot the droplet velocity at different activities. For this range of parameters, these two quantities exhibit a linear relation, which is consistent with the fact that the characteristic velocity generated by singular and non-singular distortions of the director field in active nematics is~\cite{PhysRevLett.110.228101} $v\sim \zeta R/\eta$, where $\eta$ is the absolute viscosity $\eta=\nu \rho$. As illustrated in Fig.~\ref{fig:directIncl_Velocity}(b) and (c), the droplet becomes more asymmetric as the activity increases while the distortions in the director field increase. As discussed in the main text, the droplet moves in the direction of the positive charge density, which can be seen for oblique anchoring in Fig.~\ref{fig:directIncl_Velocity}(d) and (e).

\subsection*{Zero anchoring}
\label{sec:sec:zeroAnchoring}

\begin{figure}[htb]
\centering
\includegraphics[width = 0.9\linewidth]{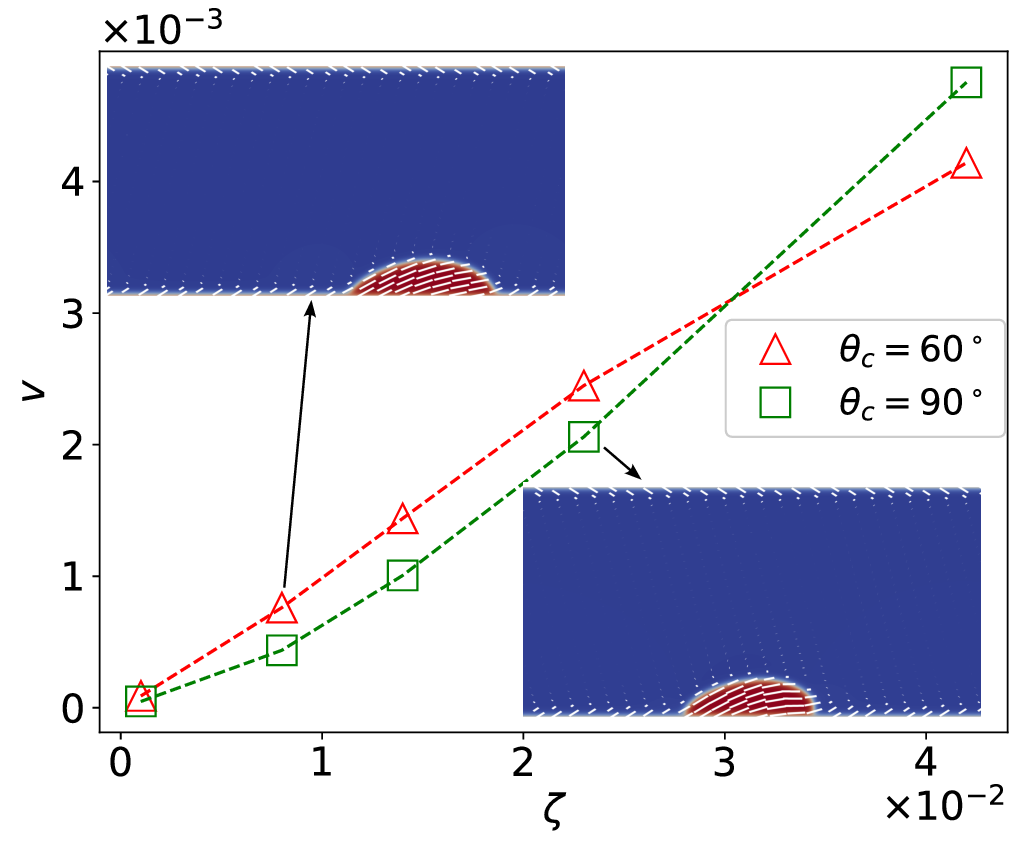}
\caption{Absolute velocity as a function of the activity, $\zeta$, for active nematics droplets on surfaces with equilibrium contact angle \(\theta_c\) and zero anchoring. The insets are snapshots of the director field in the droplet for two simulations in the steady state. In both cases, the droplet moves to the right, but droplet motion to the left was also observed (not shown).}
\label{fig:zeroAnchoring_Velocity}
\end{figure}

Finally, we have set the surface anchoring to zero. This is achieved by imposing zero gradient of the director field and $S=S_N$ at the solid-liquid interface. The initial conditions are as in the other simulations with homeotropic anchoring: uniform directors aligned vertically and velocity set to zero. We find that the director field breaks the left-right symmetry and the droplet moves at lower values of the activity. This is different from the behaviour of the suspended droplets reported in Ref.~\cite{PhysRevLett.112.147802} due to differences in the model (imposed thermodynamic interfacial anchoring) and no-slip boundary conditions at the substrate.  In Fig.~\ref{fig:zeroAnchoring_Velocity}, the droplet velocity is plotted as a function of the activity. The insets of Fig.~\ref{fig:zeroAnchoring_Velocity} depict the moving droplets at zero anchoring. We find that the droplet shape becomes asymmetric and the director field oblique as the symmetric vertical configuration is unstable to small perturbations. This is similar to the results for oblique surface anchoring. At higher activities, the droplet evaporates as before.

\subsection*{Criteria to classify the regimes with homeotropic anchoring}

In order to classify the three main regimes for the homeotropic surface, two parameters were used: the order of magnitude of the standard deviation of the linear fit used to determine the droplet velocity (droplet position against time), \(\mathcal{O}(\sigma_m)\), and the coefficient of the linear fit, \(r^2\). The order of magnitude was calculated as \(\mathcal{O}(\sigma_m) = \log(\sigma_m)\), while the position was calculated by locating the droplet's leftmost interface over time, as close to the surface as possible. \(r^2\) indicates how well the data fits the linear fit of the position, which was higher for droplets moving at a constant velocity, while \(\sigma_m\) indicates how scattered the actual data points are around the linear fit, which is higher for chaotic droplets, whose movement is not steady.

If \(\sigma_m \approx 0\), the droplet is static. This is to be expected, since the position of static droplets is always the same. The droplet is in the linear state if \(r^2 \geq 0.9\), since droplets in the linear state are those whose motion gives the best linear fit. Finally, if \(\mathcal{O}(\sigma_m) \geq -5\), the droplet is in the chaotic state. Since their motion, which oscillates, often has a linear component, the scattering is much more noticeable in this regime, as opposed to the other droplets. Visual inspection of a few cases confirm that the algorithm is reliable.

To identify the three remaining regimes, different algorithms were used for each case.

For division, a ``burn'' algorithm was applied. The algorithm consists in the following: (1) the first point where \(\phi \geq 0\) (i.e.: where the concentration of the nematic is greater than the concentration of the isotropic fluid), ignoring the surface, is found and its value is changed to 2 (``burning''), (2) all squares where \(\phi \geq 0\) that are adjacent to ``burning'' squares (excluding diagonals) have their value changed to 3 (``will burn''), (3) the ``burning'' squares are changed to 4 (``burnt''), (4) the squares that ``will burn'' are changed to ``burning'', (5) steps 2-4 are repeated until there are no more ``burning'' squares, (6) if there are still squares where \(\phi \geq 0\), the droplet has split into at least two pieces. This algorithm was implemented at every time interval, since broken droplets often merge, albeit temporarily.

Detached droplets were determined by checking if there was at least one full ``line'' of nematic between the first point in the nematic and the surface. Likewise for division, this was performed at every time step, since detached droplets can temporarily return to the surface.

Finally, evaporated droplets are found by checking if \(\phi \leq 0\) across the entire system, excluding the surface. This check was made at the last time interval, since, similarly to wetting layers, evaporated droplets stay evaporated. Although the value of $\phi$ changes with activity in the isotropic component, it remains smaller than $0$.

%%%END OF MAIN TEXT%%%



%%%REFERENCES%%%
\bibliography{rsc} 

\end{document}